\documentclass[prd,superscriptaddress,nofootinbib,floatfix, 11pt]{revtex4-1}
\usepackage[colorlinks=true,
 linkcolor=blue,
urlcolor=magenta,
citecolor=blue]{hyperref}
\usepackage[utf8]{inputenc}
\usepackage{xcolor}
\usepackage{multirow}
\usepackage[a4paper, left=2.5cm, right=2.5cm, top=2.5cm, bottom=2.5cm]{geometry}
\usepackage[T1]{fontenc}
\usepackage{amsmath,amssymb,amsthm}  
\usepackage{amsfonts}
\usepackage{mathrsfs}  
\usepackage{latexsym,bm}
\usepackage{graphicx}      
\usepackage{wrapfig}    
\usepackage{indentfirst}
\usepackage{slashed}  
\usepackage{units} 
\usepackage{fancybox} 
\usepackage{array}    
\usepackage{tabularx}    
\usepackage{booktabs}    
\usepackage{subcaption}
\usepackage{makecell}

\usepackage{caption}
\usepackage{ragged2e}

\newcommand{\beq}{\begin{eqnarray}}
\newcommand{\eeq}{\end{eqnarray}}
\newcommand{\ben}{\begin{enumerate}}
\newcommand{\een}{\end{enumerate}}
\newcommand{\non}{\nonumber\\ }

\newcommand{\xsl}{ x \hspace{-2.2truemm}/ }

\allowdisplaybreaks[4]

\begin{document}

\title{\LARGE Semileptonic decays $D_{(s)} \to \eta^{(\prime)} \ell^+ \nu_\ell$ from QCD Light-Cone Sum Rules}

\author{Xiao-En Huang}
\affiliation{School of Physics and Electronics, Hunan University, 410082 Changsha, China}

\author{Shan Cheng}\email[Corresponding author: ]{scheng@hnu.edu.cn}
\affiliation{School of Physics and Electronics, Hunan University, 410082 Changsha, China}
\affiliation{School for Theoretical Physics, Hunan University, 410082 Changsha, China}
\affiliation{Hunan Provincial Key Laboratory of High-Energy Scale Physics and Applications, 410082 Changsha, China}

\author{De-Liang Yao}\email[Corresponding author: ]{yaodeliang@hnu.edu.cn}
\affiliation{School of Physics and Electronics, Hunan University, 410082 Changsha, China}
\affiliation{School for Theoretical Physics, Hunan University, 410082 Changsha, China}
\affiliation{Hunan Provincial Key Laboratory of High-Energy Scale Physics and Applications, 410082 Changsha, China}

\date{\today} 
	
\begin{abstract}

In light of recent precision measurements from BESIII, we reanalyze the $D, D_s \to \eta^{(\prime)}$ transition form factors using QCD light-cone sum rules, incorporating high-twist contributions and well-established next-to-leading-order QCD corrections. Our analysis confirms the chiral enhancement effect arising from twist-3 light-cone distribution amplitudes of the pseudoscalar mesons, and demonstrates a rapid convergence of the operator product expansion. The resulting high-accuracy form factors enable us to determine the optimal $\eta$-$\eta^\prime$ mixing parameters from the precise experimental data for the $D, D_s \to \eta^{(\prime)} \ell^+ \nu_\ell$ (with $\ell = e, \mu$) differential decay rates. We find that the BESIII data strongly favor a set of mixing parameters, characterized by small decay constants and a large mixing angle, in the quark flavor basis. Notably, the light-cone-sum-rule predictions for the decays $D \to \eta^{(\prime)} \ell^+ \nu_\ell$, induced by weak $c\to d$ current, reach a precision comparable to the BESIII experimental results. Nevertheless, further refined measurements and more accurate form-factor determinations will be essential to scrutinize the potential role of gluonic components in charmed meson semileptonic decays.

\end{abstract}
	
\maketitle

\newpage

\section{Introduction}\label{sec:introduction}

Charmed meson semileptonic decays play a pivotal role in hadron physics~\cite{BESIII:2020nme,Friday:2025gpj}. Charged-current transitions such as $D \to K \ell^+ \nu_\ell$ ($\ell = e, \mu$) enable precise determinations of the Cabibbo-Kobayashi-Maskawa (CKM) matrix elements~\cite{BESIII:2021bdp, BESIII:2023cym,BESIII:2024dvk,BESIII:2018ccy,BESIII:2024slx,Yao:2018tqn,Parrott:2022rgu,FermilabLattice:2022gku}, while flavor-changing neutral current processes such as $D \to \pi \ell^+\ell^-, \pi\pi \ell^+\ell^-$ are sensitive to new physics beyond the standard model (SM)~\cite{Bharucha:2020eup,
Bansal:2025hcf,LHCb:2024ely,BESIII:2024nrw}. Paralleling the anomalies observed in $B$-meson decays~\cite{Belle:2019rba,LHCb:2023zxo, HeavyFlavorAveragingGroupHFLAV:2024ctg,Gambino:2020jvv}, those in charmed meson semileptonic decays provide a complementary probe of new physics by investigating the up-type quark sector. Besides, these decays, benefiting from the clear separation of their leptonic and hadronic parts, offer a clean laboratory to reveal the inner structure of light mesons whose nature is still debated. For example, the scalar meson $f_0$ (see e.g., Refs.~\cite{Pelaez:2015qba,Yao:2020bxx} for reviews), traditionally studied in cascade decays like $D_s \to (f_0 \to \pi\pi) e^+ \nu_e$~\cite{BESIII:2023wgr, Cheng:2023knr,Hu:2025ool}, is now being supplemented by a direct analysis of the $S$-wave channel $D_s \to \left[ \pi\pi \right]_{\rm S} e^+\nu_e$ with two-pion distribution amplitudes (DAs)~\cite{Cheng:2025fux,Cheng:2025hxe}. The complex nature of the isoscalar pseudoscalar mesons ($\eta, \eta^\prime$) is also under scrutiny, which is crucial for a deeper understanding of the low-energy manifestation of QCD $U_A(1)$ anomaly and the underlying mixing mechanism~\cite{Gan:2020aco}.

In the exact quark-flavor SU(3) limit, the $\eta^\prime$ meson is a flavor-singlet state, while the $\eta$ meson belongs to the flavor-octet representation. However, large SU(3)-breaking effect renders the study of their structure non-trivial. The physical states $\eta$, $\eta^\prime$ are conventionally expressed as the admixture of the flavor-singlet and -octet states $\eta_1$, $\eta_8$, which is called singlet-octet (SO) mixing scheme. The SO mixing scheme was initially formulated to study low-energy $\eta,\eta^\prime$ production and decay processes in chiral perturbative theory; See, e.g., Refs.~\cite{Witten:1978bc,Leutwyler:1997yr} and references therein. Two mixing angles are usually required in the SO mixing scheme, owing to the large SU(3)-symmetry breaking effect. In 1998, an alternative mixing scenario was proposed in Ref.~\cite{Feldmann:1998vh}, in which the physical states $\eta, \eta^\prime$ are expressed as linear combinations of orthogonal states $\eta_q$, $\eta_s$, requiring only a single mixing angle. Hereafter, we refer to it as the quark-flavor (QF) mixing scheme. The QF scheme offers a complementary method for studying isoscalar pseudoscalar eta mesons, especially in the hard reactions where high-precision measurement is feasible. It has been successfully applied to diverse phenomena in $\eta$-$\eta^\prime$ physics. Key applications include the study of electromagnetic form factors (FFs) in perturbative QCD~\cite{Feldmann:1997vc,Chai:2025xuz}, and the analysis of semileptonic heavy meson decays using various theoretical methods based on the factorization hypothesis~\cite{Ball:1995zv,Colangelo:2001cv,Azizi:2010zj,Offen:2013nma,Ball:2007hb,Aliev:2002tra,Duplancic:2015zna,Zhang:2025yeu}. 

In this work, we investigate the semileptonic decays $D, D_s \to \eta^{(\prime)} l^+ \nu_l$ within the framework of QCD light-cone sum rules (LCSRs) using the QF mixing scheme. The nonperturbative quark-gluon dynamics of the involved hadrons is embodied into their light-cone DAs (LCDAs)~\cite{Balitsky:1987bk}, which are ordered by twist. In recent years, significant progresses have been made on the light-meson LCDAs from lattice QCD (LQCD) \cite{Bali:2021qem,RQCD:2022xux,RQCD:2019osh,Cloet:2024vbv,Detmold:2025lyb} and data-driven analyses \cite{Cheng:2020vwr,Agaev:2012tm}. 
To date, analyses of $\pi$ and $K$ meson LCDAs have been developed to twist-4 accuracy~\citep{Ball:1998je,
Braun:1999uj,Bijnens:2002mg,Ball:2006wn}. As for the eta mesons,  the LCDAs for their QF states, $\eta_q$ and $\eta_s$, can be defined analogously to the pion. 
Consequently, they admit a similar expansion in Gegenbauer polynomials, where the distinct expansion coefficients reflect the SU(3) flavor-breaking effects. Building upon earlier LCSR studies \cite{Ball:2007hb,Azizi:2010zj,Offen:2013nma}, we obtain the $D, D_s \to \eta^{(\prime)}$ transition FFs, fully incorporating the well-established next-to-leading-order (NLO) QCD corrections. Specifically, the derived LCSR FFs include the NLO hard functions for both the leading-twist and two-particle twist-3 LCDAs, the NLO soft function arising from the three-particle configuration, as well as the leading two-gluon contribution. In addition, the twist-5 and 6 contributions are also estimated through a convolution of the leading-twist LCDAs with the vacuum condensate density~\cite{Rusov:2017chr}. 

The obtained $D, D_s \to \eta^{(\prime)}$ transition FFs are used to determine the optimal QF mixing parameters. Four different sets of mixing parameters, i.e., the decay constants and the mixing angle, are tested by confronting our results to the recent high-precision measurements from the BESIII collaboration~\cite{BESIII:2025hjc,BESIII:2023gbn,BESIII:2024njj}.  We find that the BESIII data strongly favor a set of mixing parameters~\cite{Bali:2021qem}, characterized by small decay constants $f_{\eta_q} = \left( 1.02^{+0.02}_{-0.05} \right) f_\pi$ and $f_{\eta_s} = \left( 1.37^{+0.04}_{-0.06} \right) f_\pi$, along with a larger mixing angle $\phi = 39.6^{+1.2}_{-2.1}$ degrees, where $f_\pi = 130$ MeV. With this set of parameters, we predict the transition FFs at zero momentum transfer $q^2=0$ , which are inputs for the extraction of the CKM matrix elements $V_{cd}$ and $V_{cs}$. It is also found that the operator product expansion (OPE) series exhibits good convergence through twist-3 and higher. Since the LCSR FFs are only valid in the range of low momentum transfers, $0 \leqslant \vert q^2 \vert \lesssim 0.4$ GeV$^2$, we extrapolate them to the whole kinematical region with the help of the Bourrely-Caprini-Lellouch (BCL) parametrization~\cite{Bourrely:2008za}. Based on the FFs, we further make predictions for the differential decay rate and branching ratios. Good agreement between our results and the BESIII measurements~\cite{BESIII:2025hjc,BESIII:2023gbn,BESIII:2024njj} is observed for the decays $D \to \eta \ell \nu_\ell$, $D_s \to \eta \ell \nu_\ell$, and $D_s \to \eta^\prime \ell \nu_\ell$, except for $D \to \eta^\prime \ell \nu_\ell$. For the $D \to \eta^\prime \ell^+ \nu_\ell$ decay, the predicted differential decay rate shows a slight tension with the BESIII result, particularly in the intermediate-to-high momentum transfer region.

The manuscript is structured as follows. The $\eta$-$\eta^\prime$ mixing schemes and the LCSR calculation of the $D,D_s \to \eta^{(\prime)}$ transition FFs are detailed in sections~\ref{sec.mixing} and~\ref{sec:LCSRs-FFs}, respectively. 
Section~\ref{sec:semileptonic} presents the numerical results of the FFs, the differential decay widths and branching ratios. Comparison with the BESIII measurements and other theoretical determinations is also discussed. Section~\ref{sec:conclu} contains our summary and outlook. Explicit expressions of the $D \to \eta_q, \eta_g$ FFs and LCDAs of light pseudoscalar mesons are relegated to Appendices~\ref{app:FF-piece} and~\ref{app:LCDAs}, respectively.

\section{Mixing schemes\label{sec.mixing}}

We begin by introducing the $\eta$-$\eta'$ mixing scheme, which is essential for the subsequent calculation of the $D, D_s \to \eta^{(\prime)}$ form factors.  In the SO scheme, the $\eta$ and $\eta'$ mesons are expressed in terms of the SU(3) flavor-octet and flavor-singlet states $\eta_8$, $\eta_1$ via~\cite{Leutwyler:1997yr} 
\beq \left( \begin{array}{cc} \vert \eta \rangle \\ \vert \eta^{'} \rangle \end{array}
\right) = U(\theta)
\left( \begin{array}{cc} \vert \eta_8 \rangle = \frac{1}{\sqrt{6}} \vert {\bar u}u + {\bar d}d - 2 {\bar s}s \rangle \\ \vert \eta_1 \rangle = \frac{1}{\sqrt{3}} \vert {\bar u}u + {\bar d}d + {\bar s}s \rangle \end{array} \right)\ ,\quad U(\theta)\equiv\left(\begin{array}{cc} \cos \theta_8 & -\sin \theta_1 \\ \sin\theta_8 & \cos \theta_1 
\end{array} \right) \ .
\label{eq:flavor18states}\eeq
The sine and cosine functions in the mixing matrix represent the probability amplitudes to find the $\eta$, $\eta'$ mesons in the flavor SO basis. Only in the exact SU(3) flavor symmetric case, the two angles are identical $\theta_1=\theta_8\equiv\theta$. In reality, the significant breaking effect of SU(3) flavor symmetry, which can reach 10–20\%~\cite{Leutwyler:1997yr}, necessitates the use of two distinct mixing angles, $\theta_1$ and $\theta_8$. 

A more convenient mixing scheme for QCD LCSR calculation is the QF scheme, proposed by Feldmann, Kroll, and Stech in Refs.~\cite{Feldmann:1998sh,Feldmann:1999uf}. In this scheme, the $\eta^{(\prime)}$ mesons are described by the orthogonal quark flavor basis ${\eta_q, \eta_s}$ as
\beq \left( \begin{array}{cc} 
\vert \eta \rangle \\ \vert \eta^{'} \rangle
\end{array} \right) =
U(\phi)
\left( \begin{array}{cc}
\vert \eta_{q} \rangle = \frac{1}{\sqrt{2}} \vert {\bar u} u + {\bar d}d \rangle \\
\vert \eta_s \rangle = \vert {\bar s}s \rangle
\end{array} \right)\ ,\quad U(\phi)\equiv\left( \begin{array}{cc}
\cos \phi & -\sin \phi \\ \sin\phi & \cos \phi  
\end{array} \right)\ .
\label{eq:SO-QF} \eeq
It is feasible to introduce just one mixing angle $\phi$, since the Okubo–Zweig–Iizuka (OZI) rule violating effect is negligible~\cite{Feldmann:1998vh}. The angle $\phi$ in the flavor basis is related to the singlet-octet mixing angle $\theta$ by 
\begin{align}
    U(\phi)= U(\theta) U(\theta^\prime)\ ,\quad U(\theta^\prime) = \left( \begin{array}{cc}
\cos \theta^\prime & -\sin \theta^\prime \\ \sin\theta^\prime & \cos \theta^\prime  
\end{array} \right)\ ,
\end{align}
where $U(\theta^\prime)$ denotes the mixing matrix between the QF and the SO bases, with $\cos \theta^\prime = 1/\sqrt{3}$ and $\sin\theta^\prime = \sqrt{2/3}$.

The relevant decay constants are defined by the matrix elements of axial-vector currents $A^{q}_{\mu} = \frac{1}{\sqrt{2}} \left( {\bar u} \gamma_\mu\gamma_5 u + {\bar d} \gamma_\mu\gamma_5 d \right)$ and $A_\mu^{s} = {\bar s} \gamma_\mu\gamma_5 s$. For the physical eigenstates ($\eta,\eta^\prime$) and the flavor eigenstates ($\eta_q, \eta_s$), the decay constants read
\beq 
\langle 0 \vert A_\mu^r \vert \eta^{(\prime)}(p) \rangle = i f_{\eta^{(\prime)}}^r p_\mu\ , \quad
\langle 0 \vert A_\mu^r \vert \eta_{t}(p) \rangle = i f_{\eta_r}\delta^{rt} p_\mu\ ,\quad r,t\in\{q,s\}\ , \label{eq:decaycons}
\eeq
respectively. They are related through the same mixing matrix $U(\phi)$ as that for the states~\eqref{eq:SO-QF}: 
\beq \label{eq.decay.consants.eta.f}
\left( \begin{array}{cc}
f_{\eta}^q & f_{\eta}^s\\ f_{\eta^{'}}^q & f_{\eta^{'}}^s \end{array} \right)
= U(\phi)
\left( \begin{array}{cc} f_{\eta_q} & 0 \\ 0 & f_{\eta_s} \end{array} \right). \eeq
Furthermore, the masses of the flavor eigenstates can be expressed in terms of the pion and kaon masses via the relations, 
\beq &&
m_{qq}^2 = \frac{\sqrt{2}}{f_{\eta_q}} \langle 0 \vert m_u {\bar u} i \gamma_5 u + m_d {\bar d} i \gamma_5 d \vert \eta_q \rangle = m_{\pi}^2\ , \non
&&m_{ss}^2 = \frac{2}{f_{\eta_s}} {\langle 0|} m_s {\bar s} i \gamma_5 s \vert \eta_s \rangle = 2 m_K^2-m_{\pi}^2\ .\label{eq.mass.qq.ss}
\eeq
Likewise, for the singlet and octet states $\eta_1$ and $\eta_8$, the decay constants are defined by 
\begin{align}
    \langle 0 \vert A_\mu^i \vert \eta^{(\prime)}(p) \rangle = i f_{\eta^{(\prime)}}^i p_\mu\ , \quad
\langle 0 \vert A_\mu^i \vert \eta_{j}(p) \rangle = i f_{\eta_r}\delta^{ij} p_\mu\ ,\quad i,j\in\{1,8\}\ , \label{eq:decaycons2}
\end{align}
with the axial-vector currents $A_\mu^1=\frac{1}{\sqrt{3}}(\bar{u}u+\bar{d}d+\bar{s}s)$ and $A_\mu^8=\frac{1}{\sqrt{6}}(\bar{u}u+\bar{d}d-2\bar{s}s)$. The singlet and octet decay constants are related to their QF basis counterparts by a pure rotation,
\beq
\left( \begin{array}{cc}
f_{\eta}^8 & f_{\eta}^1\\ f_{\eta^{'}}^8 & f_{\eta^{'}}^1 \end{array} \right)
= U(\phi)
\left( \begin{array}{cc} f_{\eta_q} & 0 \\ 0 & f_{\eta_s} \end{array} \right)U^\dagger(\theta^\prime)\ . 
\eeq
For instance, for the singlet decay constants, we have
\beq 
f_{\eta}^1=\sqrt{\frac{2}{3}}\cos \phi f_{\eta_q}-\sqrt{\frac{1}{3}}\sin \phi f_{\eta_s},\qquad
f_{\eta^{\prime}}^1=\sqrt{\frac{2}{3}}\sin \phi f_{\eta_q}+\sqrt{\frac{1}{3}}\cos \phi f_{\eta_s}\ . \label{eq:decaycons-eta1} 
\eeq 
Note that the above mixing scheme may be extended to include purely gluonic contributions, as detailed in Refs.~\cite{Feldmann:1998sh,Cheng:2008ss}.

\section{Form factors from LCSRs}\label{sec:LCSRs-FFs}

In QF basis, the derivation of the $D \to \eta_q$ FFs using LCSRs begins with a correlation function of the bilocal currents $J_\mu = {\bar d} \gamma_{\mu} c$ and $J_{5} = m_c {\bar c} i \gamma_5 d$, sandwiched between the vacuum and the flavour eigen state $\eta_q$, 
\beq
F^{\eta_q}_{\mu}(p,q) &=& i\int d^4xe^{iq\cdot x} 
\langle \eta_q(p) \vert T \{ J_\mu(x) J_5(0) \} \vert 0 \rangle  \label{eq.correlation.fun}
\eeq
Here, $p_\mu$ is the four-momentum of the $\eta_q$ meson, and $q_\mu$ is the momentum transferred by the weak vector current $J_\mu$. 
The correlation function can be further decomposed into the form
\beq
F^{\eta_q}_{\mu}(p,q)&=& F^{\eta_q}(q^2,{\tilde q}^2) \,p_{\mu} + \tilde{F}^{\eta_q}(q^2,{\tilde q}^2)\,q_{\mu}\ , \label{eq:correlator}
\eeq
where $F^{\eta_q}$ and ${\tilde F}^{\eta_q}$ are Lorentz-invariant functions with respect to the invariant mass squares $q^2$ and ${\tilde q}^2 \equiv (p+q)^2$. Likewise, for the $D_s \to \eta_s$ transition FFs, the corresponding correlation can be obtained by replacing the down quark with the strange quark. That is, the weak and interpolating currents in Eq.~\eqref{eq.correlation.fun} should be changed to $J_\mu = {\bar s} \gamma_{\mu} c$ and $J_{5} = m_c {\bar c} i \gamma_5 s$, respectively.

The correlation functions can be formulated in two ways. On the one hand, in the spacelike region of negative momentum transfer squared $q^2 <0 $ and the decay region $0 \leq q^2 < (m_{D} - m_{\eta_q})^2$, it can be calculated directly in QCD at the quark–gluon level by applying the OPE. On the other hand, in the physical region of ${\tilde q}^2$, the long-distance quark–gluon interaction between bilocal currents begins to produce hadrons, known as hadronization. In this case, the correlation function can be expressed as a hadronic dispersive representation, with the spectral density being the sum of contributions from all possible intermediate states.
 
More specifically, the OPE is valid when the final state $\eta_q$ is energetic, so the QCD calculation of the correlation function is restricted to not too large momentum transfer squared $0 \leq \vert q^2 \vert \leq q^2_{\rm max} \simeq m_c^2 - 2 m_c \chi$, with $\chi \sim 500$ MeV being a typical hadronic scale parameter. For the spacelike region $q^2 < 0$ considered here, the light-cone OPE works even better than for small positive $q^2$.  
Provided that $q^2, {\tilde q}^2 \ll m_c^2$, the charm quark field emitted and aborbed by the currents is highly virtual and propagates at small $x^2$. Consequently, the charm quark propagator near the light cone $x^2 \sim 0$ can be expanded as
\beq 
S_c(x,0)&& \equiv -i \langle 0 \vert T \{ c_i(x), \bar{c}_j(0) \} \vert 0\rangle = \frac{-i m_c^2}{4\pi^2} \bigg[ \frac{K_1(m_c\sqrt{\vert x^2 \vert})}{\sqrt{\vert x^2 \vert}} + \frac{i \xsl K_2(m_c \sqrt{\vert x^2 \vert})}{\vert x^2 \vert} \bigg] \delta_{ij} \non 
&&- \frac{ig_s}{16\pi^2} \int_0^1 dv \bigg[ m_c K_0(m_c\sqrt{\vert x^2 \vert}) (\mathbf{G} \cdot \sigma) 
+ \frac{im_c K_1(m_c\sqrt{\vert x^2 \vert})}{\sqrt{\vert x^2 \vert}} \Big( \bar{v} \xsl (\mathbf{G} \cdot \sigma) + v (\mathbf{G} \cdot \sigma) \xsl \Big)  \non 
&&\left. -2v\bar{v} \bigg( im_cK_0(m_c\sqrt{\vert x^2 \vert}) - \frac{m_c\slashed{x}}{\sqrt{\vert x^2 \vert}}K_1(m_c\sqrt{\vert x^2 \vert}) \bigg) x_{\nu}D_{\mu}\mathbf{G}^{\mu \nu}(vx) \right. \non 
&&\left. +K_0(m_c\sqrt{\vert x^2 \vert})(2v\bar{v}-1)\gamma_{\nu}D_{\mu}\mathbf{G}^{\mu \nu}(vx) -v\bar{v}(1-2v)K_0(m_c\sqrt{\vert x^2 \vert})x_\nu \slashed{D}D_\mu \mathbf{G}^{\mu \nu}(vx)\right. \non 
&&\left. +iv\bar{v}K_0(m_c\sqrt{\vert x^2 \vert})\epsilon_{\sigma \mu \nu \rho }x^\sigma \gamma^\mu \gamma_5 D^\nu D_\alpha \mathbf{G}^{\alpha \rho}(vx)\right. \non 
&&\left. +v\bar{v}\sqrt{\vert x^2 \vert}K_1(m_c\sqrt{\vert x^2 \vert})\sigma_\rho ^{\nu}D_\nu D_\mu \mathbf{G}^{\mu \rho}(vx) \right.\bigg] \delta_{ij}\ . \label{eq:propagator-c} \eeq
Here, $\mathbf{G}^{\mu\nu}(vx) = \mathbf{G}^{\mu\nu a}(vx) {\lambda^a}/{2}$ is the field strength tensor, while the antisymmetric Dirac tensor is defined as $\sigma^{\mu\nu} = \frac{i}{2} [\gamma^\mu, \gamma^\nu]$. The function $K_i$ denotes the modified Bessel function of the second kind and $D_{\mu}$ is the covariant derivative. The first and second terms in Eq.~(\ref{eq:propagator-c}) correspond to the free charm-quark propagator and the leading quark-gluon interactions, respectively. The incorporation of higher-twist effects is achieved through the inclusion of terms containing covariant derivatives~\cite{Rusov:2017chr}.

The OPE expresses the invariant amplitudes $F^{\eta_{q}}$ and ${\tilde F}^{\eta_{q}}$ as twist-ordered convolutions of hard-scattering functions with LCDAs.  
We use the amplitude $F$ as an illustrative example to demonstrate the factorization formalism: 
\beq F^{\eta_q}(q^2, {\tilde q}^2) &=& \sum_{t=2}^6 \int_0^1 du \, T_{q{\bar q}}^{(t), {\rm LO}}(u,q^2,{\tilde q}^2) \, \phi^{(t)}_{q{\bar q}}(u) 
+\sum_{t=2}^3 \int_0^1 du \, T_{q{\bar q}}^{(t), {\rm NLO}}(u,q^2,{\tilde q}^2) \, \phi^{(t)}_{q{\bar q}}(u) \non 
&+& \sum_{t=3}^4 \int{\cal D} \alpha_i \, T_{q{\bar q}g}^{(t)}(u, \alpha_i, q^2, {\tilde q}^2) \, \phi_{q{\bar q}g}^{(t)}(\alpha_i)\ , 
\label{eq:CF-flavor-OPE} 
\eeq
where summation over twists is implied. The hard-scattering amplitudes $T$ capture the physics in the large recoil region of $q^2$, i.e., $q^2 \to 0$. The light-quark LCDAs of the $\eta_{q}$ mesons are defined by the matrix element of a nonlocal operator, sandwiched between the vacuum and the on-shell $\eta_{q}$ state. The above expression includes contributions from various Fock states, described by their respective LCDAs: the two-particle ($\phi_{q{\bar q}}$) and three-particle ($\phi_{q{\bar q}g}$). 
For easy reference, we collect the LCDAs for the flavor eigenstate $\eta_q$ in Appendix~\ref{app:LCDAs}, together with the ones for $\eta_s$ and $\eta_{g}$ (two gluonic component). These eigenstates all contribute to the physical $\eta$ and $\eta^\prime$ states, with their respective probability amplitudes determined at leading order (LO) and NLO in literature~\cite{Ball:2006wn,Ball:2007hb}. 
For the NLO corrections, the two-particle twist-2 and twist-3 amplitudes are derived in the state-of-the-art QCD framework~\citep{Duplancic:2008ix}.

To saturate the correlation function in the physical region, we insert a complete set of intermediate hadronic states, with the momentum $\tilde{q}$ and quantum numbers of the $D$ meson, between the bilocal currents. This yields a hadronic dispersion relation encompassing the contribution from the $D$ meson, its excited states and the non-resonant continuum spectral. Namely, 
\beq 
&&F^{\eta_q}(q^2,{\tilde q}^2) = \frac{2m_{D}^2 f_{D} \, f^{D \to \eta_q}_ +(q^2)}{m_{D_s}^2-{\tilde q}^2} + \frac{1}{\pi} \int_{s_0}^{\infty}ds
\frac{\rho^{D \to \eta_q}}{s-{\tilde q}^2}\ , \non
&&\tilde{F}^{\eta_q}(q^2,{\tilde q}^2) = \frac{m_{D}^2f_{D} \left[ f^{D \to \eta_q}_+(q^2) + f^{D \to \eta_q}_-(q^2) \right]}{m_{D}^2-{\tilde q}^2} + \frac{1}{\pi} \int_{s_0}^{\infty}ds
\frac{{\tilde \rho}^{D \to \eta_q}}{s-{\tilde q}^2}\ .
\label{eq:CF-F-hadron} 
\eeq
The ground-state contribution is isolated through the introduction of a continuum threshold $s_0$, with the remaining spectral strength from excited states and the continuum encoded in the spectral densities $\rho$ and ${\tilde \rho}$. This ground-state term is proportional to the decay constant $f_{D}$ and the $D \to \eta_q$ transition FFs $f_{\pm}^{D\to \eta_q}$, which are defined via the matrix elements
\beq &&\langle D(\tilde{q}) \vert {\bar c} i m_c \gamma_5 d \vert 0 \rangle = m_{D}^2 f_{D}\ , \non
&&\langle \eta_q(p) \vert {\bar d} \gamma_{\mu} c \vert D(\tilde{q}) \rangle = 2 f_+(q^2) p_\mu +  \left[ f_+(q^2) + f_-(q^2) \right] q_\mu\ , \label{eq:Ds2etap-FF} \eeq 
respectively. The superscripts of $f_{\pm}$ are suppressed for brevity and the abbreviation $f_{+-}\equiv f_++f_-$ is used hereafter. An equivalent parameterization for the latter matrix element reads
\beq \langle \eta_q(p) \vert {\bar d} \gamma_{\mu} c \vert D(\tilde{q}) \rangle = f_+(q^2) \left[ \left( 2p + q \right)_\mu -\frac{m_{D}^2 - m_{qq}^2}{q^2} q_\mu \right] + f_0(q^2) \frac{m_{D}^2 - m_{qq}^2}{q^2} q_\mu\ , \label{eq:Ds2etap-FF-1}\eeq  
where $f_+$ and $f_0$ are usually called vector and scalar FFs, respectively. The scalar FF $f_0$ is associated with $f_+$ and $f_-$ by~\cite{Charles:1998dr}
\beq 
f_0(q^2) = f_+(q^2) + f_-(q^2) \frac{q^2}{m_{D}^2 - m_{qq}^2}\ ,
\label{eq:FF-relation}
\eeq
with $m_{qq}$ specified in Eq.~\eqref{eq.mass.qq.ss}.

By applying quark-hadron duality, we relate the hadronic description to the OPE calculation, which we express via a dispersion relation in the invariant mass squared, $\tilde{q}^2$. A Borel transformation is subsequently applied to suppress unimportant contributions: from higher-twist LCDAs on the quark-gluon level, and from excited states and the continuum spectrum on the hadronic level. Finally, we obtain the $D\to \eta_q$ transition FFs
\begin{align}\label{eq.ff.etaq}
    f_{i}^{D\to \eta_q} &= \frac{f_{\eta_q}}{\sqrt{2}}\left[ \sum_{t=2}^6 \, f_{i, {\rm LO}}^{D \to \eta_q, (t)}(q^2) + \sum_{t=2}^3 \, f_{i, {\rm NLO}}^{D \to \eta_q, (t)}(q^2) \right]\ ,\quad i\in\{+,0\}\ .
\end{align}
Note that the scalar FF $f_0$ is deduced from the OPE result of the combined FF $f_{+-}$ and the vector FF $f_+$ via Eq.~\eqref{eq:FF-relation}.  The former is given by the invariant amplitude ${\tilde F}$, as can be seen from Eq.~\eqref{eq:CF-F-hadron}, which has been calculated for both the $q{\bar q}$ and $q{\bar q}g$ Fock states up to twist-4 accuracy. The state-of-the-art QCD calculation of the vector FF $f_+$ is highly complete. For the $q\bar{q}$ Fock state, the calculation includes contributions up to twist-4 at LO and extends to twist-3 at NLO. Furthermore, the $q\bar{q}g$ Fock state is included up to twist-4 at LO. To estimate higher-order effects, we adopt the methodology established for $B \to \pi$ transitions, which expresses the twist-5 and -6 contributions from the $q\bar{q}$ state as a convolution of the leading-twist LCDA with the vacuum condensate density \cite{Rusov:2017chr}. 

Analogously, one can derive the $D_s\to \eta_s$ and $D_{(s)}\to \eta_g$ transition FFs, 
\begin{align}
f_{i}^{D_s\to \eta_s} &= {f_{\eta_s}}\left[ \sum_{t=2}^6 \, f_{i, {\rm LO}}^{D_s \to \eta_s, (t)}(q^2) + \sum_{t=2}^3 \, f_{i, {\rm NLO}}^{D_s \to \eta_s, (t)}(q^2) \right]\ ,\label{eq.ff.etas}\\
f_i^{D_{(s)}\to\eta_g} &= f_{\eta^{(\prime)}}^1 f_{i, {\rm NLO}}^{D_{(s)} \to \eta_g, (t=2)}(q^2)\ ,\quad i\in\{+,0\}\ .\label{eq.ff.etag}
\end{align}
The two-gluon contribution begins at NLO, and its current evaluation is at twist-2 accuracy~\cite{Ball:2007hb,Duplancic:2015zna}, which are provided in Appendix~\ref{app:FF-piece} for easy reference.

Eventually, the physical transition FFs can be expressed in terms of those derived with flavor eigenstates. With Eqs.~\eqref{eq.ff.etaq}, \eqref{eq.ff.etas} and \eqref{eq.ff.etag}, one gets
\beq &&f_i^{D \to \eta^{(\prime)}}(q^2) = \frac{f_{\eta^{(\prime)}}^{q}}{\sqrt{2}} \left[ \sum_{t=2}^6 \, f_{i, {\rm LO}}^{D \to \eta_q, (t)}(q^2) + \sum_{t=2}^3 \, f_{i, {\rm NLO}}^{D \to \eta_q, (t)}(q^2) \right] + f_{\eta^{(\prime)}}^{1} \, f_{i, {\rm NLO}}^{D \to \eta_g, (t=2)}(q^2)\ , \\
 &&f_i^{D_s \to \eta^{(\prime)}}(q^2) = 
 f_{\eta^{(\prime)}}^{s} \left[ \sum_{t=2}^6 \, f_{i, {\rm LO}}^{D_s \to \eta_s, (t)}(q^2) + \sum_{t=2}^3 \, f_{i, {\rm NLO}}^{D_s \to \eta_s, (t)}(q^2) \right] +  f_{\eta^{(\prime)}}^{1} \, f_{i, {\rm NLO}}^{D_s \to \eta_g, (t=2)}(q^2)\  , 
\label{eq:f+-twist-order} 
\eeq 
where $i\in\{+,0\}$ and the mixing effects are encoded in the decay constants, as shown in Eqs.~\eqref{eq.decay.consants.eta.f} and~\eqref{eq:decaycons-eta1}.

\section{Semileptonic decays $D,D_s \to \eta^{(\prime)} l^+ \nu_l$}\label{sec:semileptonic} 

\subsection{Parameter setup}

A central obejective of this study is to test the $\eta$–$\eta^\prime$ mixing parameters by analyzing semileptonic decays of $D$ and $D_s$ mesons. To this end, we systematically evaluate the four popular parameter sets employed in the QF scheme, which are compiled in the upper panel of Table~\ref{tab:parameter-mixing-t2LCDAs}. The decay constants of the $\eta_q$ and $\eta_s$ flavor eigenstates are expressed in units of the isovector pion decay constant, $f_\pi = 0.130$ GeV~\cite{ParticleDataGroup:2024cfk}. For the twist-2 LCDAs of the flavor states $\eta_q$ and $\eta_s$, we adopt the same functional form as the pion LCDA, assuming SU(3) flavor symmetry. The lower panel of Table~\ref{tab:parameter-mixing-t2LCDAs} summarizes the resulting first two Gegenbauer coefficients, which are taken from the lattice QCD calculation of moments~\cite{RQCD:2019osh} and from data-driven analyses of the pion electromagnetic FFs~~\cite{Cheng:2020vwr} as well as the $\pi^0\gamma^\ast\gamma$ transition FFs~\cite{Agaev:2012tm}. It is found in Ref.~\cite{Khodjamirian:2009ys}, for the $D \to \pi$ transition FFs, the twist-2 contribution is not dominant due to the chiral enhancement of the twist-3 terms. Consequently, the primary uncertainty in the LCDAs stems from the chiral mass rather than the Gegenbauer coefficients $a_n$. We therefore employ the set I parameters for $a_n$ in our analysis, as it aligns with both established QCD sum rule results~\cite{Khodjamirian:2004ga} and the recent lattice determination based on a large momentum effective theory~\cite{LatticeParton:2022zqc}.

\begin{table}[t] 
\centering 
\caption{Upper panel: four sets of parameters used in the QF mixing scheme; Lower panel: three sets of parameters for the leading-twist LCDAs of QF eigenstates $\eta_q$ and $\eta_s$.}\label{tab:parameter-mixing-t2LCDAs} 
\setlength{\tabcolsep}{10pt}
\begin{tabular}{c | ccc} 
\hline
& $f_{\eta_q}$ & $f_{\eta_s}$ & $\phi \, ({\rm Degree})$  \\
\hline
Set A \cite{Bali:2021qem} & $(1.02^{+0.02}_{-0.05})f_{\pi}$ & $(1.37^{+0.04}_{-0.06})f_{\pi}$ & $39.6^{+1.2}_{-2.1}$ \\ 
Set B \cite{Feldmann:1998vh} & $(1.07\pm0.02)f_{\pi}$ & $(1.34\pm0.06)f_{\pi}$ & $39.3\pm1.0$ \\
Set C \cite{Escribano:2005qq} & $(1.09\pm0.03)f_{\pi}$ & $(1.66\pm0.06)f_{\pi}$ & $40.3\pm1.8$ \\
Set D \cite{Cao:2012nj} & $(1.08\pm0.04)f_{\pi}$ & $(1.25\pm0.09)f_{\pi}$ & $37.7\pm0.7$  \\
\hline
\hline
 & $a_2^{\eta_q} = a_2^{\eta_s} \, (1\,\mathrm{GeV})$ & $a_4^{\eta_q}=a_4^{\eta_s} \, (1\,\mathrm{GeV})$ \\
\hline
Set I \cite{Cheng:2020vwr} & $0.279\pm0.047$ & $0.189\pm0.060$ \\
Set II \cite{Agaev:2012tm} & $0.10$ & $0.10$ \\
Set III \cite{RQCD:2019osh} & $0.137^{+0.029}_{-0.031}$ & $\cdots$ \\
\hline
\end{tabular} 
\end{table}

The input parameters for the high-twist LCDAs of the $\eta_q$ and $\eta_s$ states, listed in Table~\ref{tab:parameters-LCDAs-t34}, are set equal to those of the pion LCDA. {The scale-dependent parameters are obtained at the energy scale $\mu=1$ GeV.} The Gell-Mann-Oakes-Renner (GMOR) relation $\langle \bar{q}q \rangle = -f_{\pi}^2\mu_{\pi}/2$ indicates that the chiral mass of pion is associated with the quark-condensate density, which contributes to the $D,D_s \to \eta_{q,s}$ transition FFs at twist-5 and twist-6 level~\cite{Rusov:2017chr}. Following Ref.~\cite{Khodjamirian:2009ys}, we adopt $\langle \bar{q}q \rangle (\mu=1 \, {\rm GeV}) = (-258^{+11}_{-9} \,\text{MeV})^3$, and fix the following chiral masses 
$\mu_\pi =\mu_{\eta_q} = \mu_{\eta_s} = 2.03^{+0.28}_{-0.21}$. The physical masses of the light Goldstone bosons are taken as $m_\pi = 0.139$ GeV, $m_K = 0.494$ GeV, $m_\eta = 0.548$ GeV, and $m_{\eta^\prime} = 0.958$ GeV, while the charmed meson masses are set to be $m_{D} = 1.87$ GeV and $m_{D_s} = 1.97$ GeV \cite{ParticleDataGroup:2024cfk}. For the decay constants, we use $f_D = 0.208$~GeV from the lattice QCD simulation by Ref.~\cite{Kuberski:2024pms} and $f_{D_s} = 0.251$ GeV from the experimental measurement by Ref.~\cite{BESIII:2021bdp}. The Gegenbauer coefficient in the two-gluon twist-2 LCDAs is taken to be $b_2^g = 0\pm 20$. The charm quark mass is set as ${\bar m_c}(m_c) = 1.27 \pm 0.01$~GeV within the $\overline{\rm MS}$ scheme. The renormalization scale is chosen as $\mu = \left( m_D^2 - m_c^2 \right)^{1/2} = 1.4$ GeV for all the scale-running parameters in the $D$ meson decays, hence at $1.5$~GeV in the $D_s$ decays \cite{Lin:2025cmn}. In addition, we use the PDG averages for the CKM matrix elements $\vert V_{cd} \vert = 0.218 \pm 0.004$ and $\vert V_{cs} \vert = 0.987 \pm 0.011$. 

\begin{table}[t] \centering\vspace{-2mm}
\caption{Input parameters for high-twist LCDAs of QF eigenstates $\eta_q$ and $\eta_s$.}
\label{tab:parameters-LCDAs-t34} 
\setlength{\tabcolsep}{10pt}
\begin{tabular}{lc | lc}
\hline
Parameters & Values ($\mu=$1\,GeV) & 
Parameters & Values ($\mu=$1\,GeV) \\
\hline
$f_{3 \eta_q} = f_{3 \eta_s} \simeq f_{3\pi}$ & $(0.0045 \pm 0.0015) \, {\rm GeV}^2$ & 
$\omega_{3 \eta_q} = \omega_{3 \eta_s} 
\simeq \omega_{3\pi}$ & $-1.5 \pm 0.7$ \\
$\lambda_{3 \eta_q} = \lambda_{3 \eta_s} 
\simeq \lambda_{3\pi}$ & 0 & 
$\omega_{4\eta_q} = \omega_{4\eta_s} 
\simeq \omega_{4\pi}$ & $0.2 \pm 0.1$ \\
$\delta_{\eta_q}^2 = \delta_{\eta_s}^2 
\simeq \delta_{\pi}^2$ & $(0.18 \pm 0.06) \, {\rm GeV}^2$ & $\mathcal{K}_{4\eta_q} = \mathcal{K}_{4\eta_s} = \simeq \mathcal{K}_{4\pi}$ & 0  \\
$\mu_{\eta_q} = \mu_{\eta_s} \simeq \mu_{\pi}$ & $2.03^{+0.28}_{-0.21}$ \, {\rm GeV} & & \\
\hline
\end{tabular} 
\end{table}

The primary sources of systematic uncertainty in LCSRs calculations are the Borel mass parameter $M^2$ and the continuum threshold $s_0$. Although not derived from first principles, they are strongly constrained by the requirement that the physical results be stable and reliable. The Borel mass parameter, $M^2$, reflects the virtuality of the internal quark propagator, scaling as $M^2 \sim \mathcal{O}(u m_{D_{(s)}}^2 + \bar{u} q^2 - u\bar{u} m_{\eta_r}^2)$. Its optimal value is determined by a compromise between two competing requirements: it must be large enough to ensure the convergence of the OPE, yet small enough to sufficiently suppress contributions from excited states and the continuum. The threshold parameter $s_0$ is typically chosen near the mass of the first excited hadron state. This choice is validated by the presence of a Borel mass plateau, where the extracted FFs $f_{i}(q^2)$ exhibits stability, i.e., ${{\rm d} \left[\ln f_{i}(q^2) \right]}/{{\rm d}(1/M^2)} = 0$,  over a range bounded by the factorization scale and the continuum threshold $\mu^2 < M^2 \leqslant s_0$. We employ the LCSR parameters $M^2=4.4 \pm 1.1$ GeV$^2$ and $s_0=7.0 \pm 0.6$ GeV$^2$ in our numeraical computation,
following Ref.~\cite{Offen:2013nma}. 

\subsection{Transition form factors}

With the parameters specified above, we are now in the position to make predictions of the $D, D_s \to \eta^{(\prime)}$ transition FFs.
Table \ref{tab:FF-twist-order} presents our results of FFs at the maximum recoil point $q^2=0$, where the vector and scalar FFs are identical, namely, $f_+(0) = f_0(0)$. Contributions from different twists (in OPE) and various orders (in QCD $\alpha_s$ expansion) are shown separately to assess the convergence property. Using Set A paremeters for the $\eta$-$\eta'$ mixing and incorporating NLO corrections from two-particle twist-2 and twist-3 LCDAs, our analysis confirms that the chiral enhancement originates primarily from the two-particle twist-3 LCDAs, while the contribution from their three-particle counterparts is negligible. The numerical results in Table~\ref{tab:FF-twist-order} demonstrate good convergence of the OPE through twist-3 and beyond. This is attributed to the strong suppression of higher-order terms: the three-particle contributions are suppressed by ${\cal O}(f_{3\eta_{q,s}}/m_c)$, the two-particle twist-4 terms by ${\cal O}(\delta^2_{\eta_{q,s}})$, and the twist-5 and twist-6 terms by ${\cal O}(\langle {\bar q}q \rangle/m_c^3)$. Furthermore, the two-gluon LCDA contribution has a central value of zero in our choice $b_2^g = 0 \pm 20$. Consequently, including NLO corrections ultimately reduces the form factors by $2\% - 3\%$. This suppression stems from a destructive interference between the two-particle twist-2 and twist-3 contributions, which is analogous to the phenomenon observed in the perturbative QCD calculation~\cite{Cheng:2014fwa,Cheng:2014gba}.
\begin{table}[t]
\centering
\caption{Results of the transition FFs at zero momentum transfer $q^2=0$, based on Set A parameters. Contributions from different twists in OPE and various orders in QCD $\alpha_s$ expansion are shown separately.}
\label{tab:FF-twist-order}
\setlength{\tabcolsep}{5pt}
\begin{tabular}{c|ccccc}
\hline
 & Twist-2 LO & Twist-2 NLO & Twist-3 LO & Twist-3 NLO & Twist-(4+5+6) LO \\ 
\hline
$f_{+,0}^{D \to \eta}(0)$ & $0.127$ & $0.025$ & $0.254$  & $-0.035$ & $-4.86 \times 10^{-4}$ \\
$f_{+,0}^{D \to \eta^\prime}(0)$ & $0.105$ & $0.021$ & $0.210$ & $-0.029$ & $-4.02 \times 10^{-4}$  \\
$f_{+,0}^{D_s \to \eta}(0)$ & $0.153$ & $0.031$ & $0.317$ & $-0.051$ & $-1.05 \times 10^{-4}$ \\
$f_{+,0}^{D_s \to \eta^\prime}(0)$ & $0.185$ & $0.037$ & $0.383$ & $-0.061$  & $-1.27 \times 10^{-4}$ \\
\hline
\end{tabular}  
\end{table}

Table~\ref{tab:FF-LCSRs-errors} provides the error budget of $f_{+,0}(0)$, based on the input parameters of Set A. The net uncertainties in the LCSR predictions are approximately $\pm 10 \%$, which are shown in the 2rd column of Table~\ref{tab:FF-LCSRs-errors}. The individual uncertainties arising from the input parameters are broken down and presented in columns 3 to 7. The primary uncertainties in the $D_{(s)} \to \eta$ transitions stem from the chiral masses of the flavor eigenstates $\eta_{q,s}$, which appear through chirally enhanced twist-3 contributions. In contrast, for the $D_{(s)} \to \eta^\prime$ transitions, the uncertainty is magnified further by the poorly constrained two-gluon LCDAs parameter $b_2^g$. This amplification occurs because of the marked hierarchy between the flavor-singlet decay constants: $f_{\eta^\prime}^1=0.148^{+0.003}_{-0.004}$ GeV and $f_\eta^1=0.018^{+0.006}_{-0.004}$\,GeV, which strongly weights the $b_2^g$ contribution in the $\eta^\prime$ case. All other parameter uncertainties are negligible by comparison.

\begin{table}[t]
\centering
\caption{Error budgets of the transition FFs at zero momentum transfer $q^2=0$, based on Set A parameters.}
\label{tab:FF-LCSRs-errors}
\setlength{\tabcolsep}{12pt}
\begin{tabular}{c|c|ccccc}
\hline
Channels & $f_{+,0}(0)$ & $M^2$ & $s_0$ & $m_0^{\eta_{q,s}}$ & $a_n^\pi$ & $b_2^g$  \\ 
\hline
${D \to \eta}$ & $0.370 ^{+0.031}_{-0.024}$ & $^{+0.003}_{-0.001}$ & $^{+0.005}_{-0.006}$  & $^{+0.029}_{-0.023}$ & $\pm 0.002$ & $\pm 0.006$ \\
$D \to \eta^\prime$ &$0.306^{+0.054}_{-0.052}$ &$^{+0.002}_{-0.001}$ &$^{+0.004}_{-0.005}$ &$^{+0.024}_{-0.019}$ &$\pm 0,002$ &$\pm 0.048$ \\
$D_s \to \eta$ &$0.449^{+0.038}_{-0.030}$ &$^{+0.011}_{-0.006}$ &$^{+0.006}_{-0.009} $&$^{+0.036}_{-0.028}$ &$\pm 0.003$ &$\pm 0.005$ \\
$D_s \to \eta^\prime$&$0.543^{+0.059}_{-0.052}$& $^{+0.013}_{-0.007}$&$^{+0.008}_{-0.011}$ &$^{+0.043}_{-0.033}$ &$\pm 0.003$ & $\pm 0.037$\\
\hline
\end{tabular} 
\end{table}

The LCSR approach provides reliable results only for low momentum transfers, $0 \leqslant \vert q^2 \vert \lesssim 0.4$ GeV$^2$. To extend this description across the full physical domain $0 \leqslant q^2 \leqslant (m_{D_{(s)}} - m_{\eta^{(\prime)}})^2$, we extrapolate the LCSR results to higher $q^2$ using the BCL parameterization~\cite{Bourrely:2008za},
\beq 
f_i(q^2) = \frac{1}{1-q^2/M_{R,i}^2} \sum_{k} \alpha^k_i \left[ z(q^2) - z(0) \right]^k\ , \label{eq:BCL} 
\eeq 
with $M_{R,i}$ denoting the masses of low-lying resonances in the $D_{(s)}\eta^{(\prime)}$ spectrum. The advantages of the BCL parameterization are twofold.
First, it employs a conformal mapping 
\begin{align}
 z(t)=\frac{\sqrt{t_+-t}-\sqrt{t_+-t_0}}{\sqrt{t_+-t}+\sqrt{t_+-t_0}}\ ,\quad
t_{\pm}\equiv (m_{D_{(s)}}-m_{\eta^{(\prime)}})^2\ ,\quad
t_0\equiv t_+(1-\sqrt{1-t_-/t_+})\ ,
\label{eq:BCL-z}    
\end{align}
which ensures rapid convergence. Second, it explicitly encodes the pole position of the lowest-lying resonance within the certain spectral function. In this work, the BCL parametrization is truncated at the second order $k\leq 2$, with values of the coefficients $\alpha_k$ given in Table~\ref{tab:BCL-masses-coefficients}. The appropriate resonances, with $J^P=1^-$ for the vector FF $f_+$ and $0^+$ for the scalar FF $f_0$, are also complied in Table~\ref{tab:BCL-masses-coefficients}.

\begin{table}[t] \centering\vspace{-2mm}
\caption{Pole masses of the lowest-lying excited states, in units of GeV, and the coefficients $\alpha_k$ ($0\leq k\leq 2$) in the BCL parameterization.}
\label{tab:BCL-masses-coefficients}
\setlength{\tabcolsep}{8pt}
\begin{tabular}{c | c c | ccc | ccc} 
\hline
\multirow{2}*{$f_i(q^2)$} & \multirow{2}*{$J^P$} & \multirow{2}*{$M_{R,i}^{c\to d}$} & 
\multicolumn{3}{c|}{$D \to \eta$} & \multicolumn{3}{c}{$D \to \eta^{\prime}$} \\ 
\cline{4-6}\cline{7-9}
& & & $\alpha_0$ & $\alpha_1$ & $\alpha_2$ & $\alpha_0$ & $\alpha_1$ & $\alpha_2$ \\ 
\hline
$f_+(q^2)$   & $1^-$ & 2.010 & 0.370 & -0.416 & -1.524 & 0.306 & -0.470 & -2.041   \\
$f_0(q^2)$ & $0^+$ & 2.343 & 0.370 & -0.198 & 0.404  & 0.306 & -0.013 & 0.259   \\
\hline\hline
\multirow{2.2}{*}{$f_i(q^2)$} & \multirow{2.2}{*}{$J^P$} & \multirow{2.2}{*}{$M_{R,i}^{c\to s}$} & \multicolumn{3}{c|}{$D_s \to \eta$} & \multicolumn{3}{c}{$D_s \to \eta^{\prime}$} \\ 
\cline{4-6}\cline{7-9}
& & & $\alpha_0$ & $\alpha_1$ & $\alpha_2$ & $\alpha_0$ & $\alpha_1$ & $\alpha_2$ \\ 
\hline
$f_+(q^2)$   & $1^-$ & 2.112 & 0.449 & 0.733 & -6.982 & 0.543 & -1.197 & -14.83   \\
$f_0(q^2)$ & $0^+$ & 2.317 & 0.449 & -0.262 & -6.480  & 0.543 & -0.118 & -15.39   \\ 
\hline
\end{tabular} 
\end{table}

\begin{figure}[t] \centering 
\begin{subfigure}[b]{0.49\textwidth}\centering
\includegraphics[width=\textwidth]{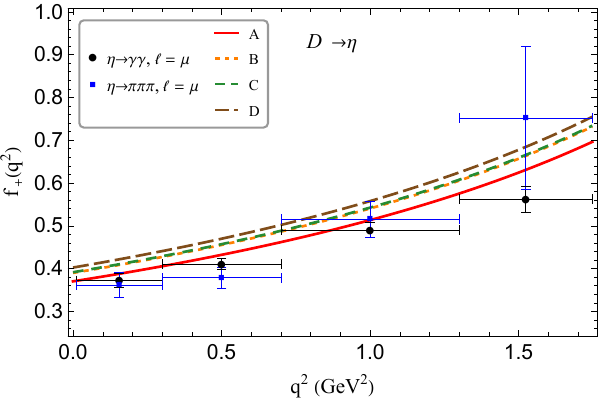}
\end{subfigure}\hfill 
\begin{subfigure}[b]{0.49\textwidth}\centering
\includegraphics[width=\textwidth]{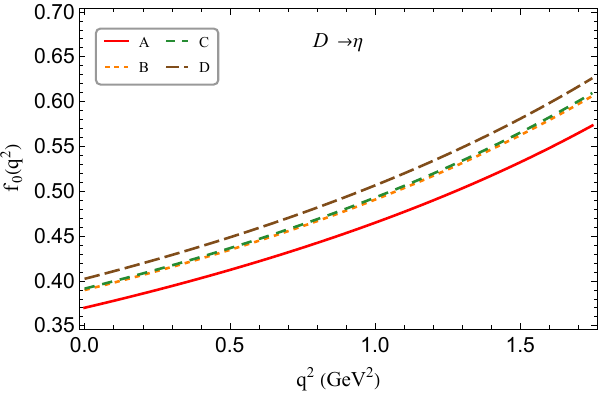}  
\end{subfigure} \non\vspace{4mm}
\begin{subfigure}[b]{0.49\textwidth}\centering
\includegraphics[width=\textwidth]{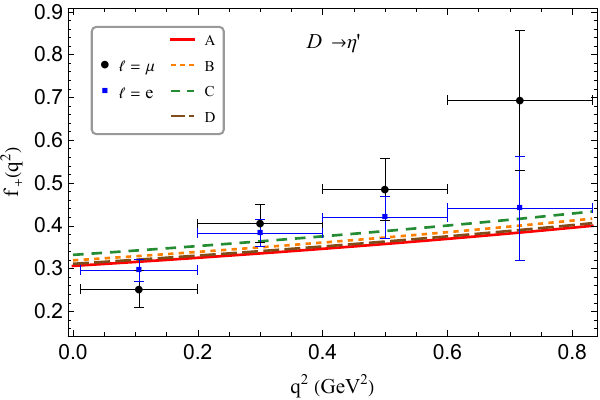} 
\end{subfigure}\hfill  
\begin{subfigure}[b]{0.49\textwidth}\centering
\includegraphics[width=\textwidth]{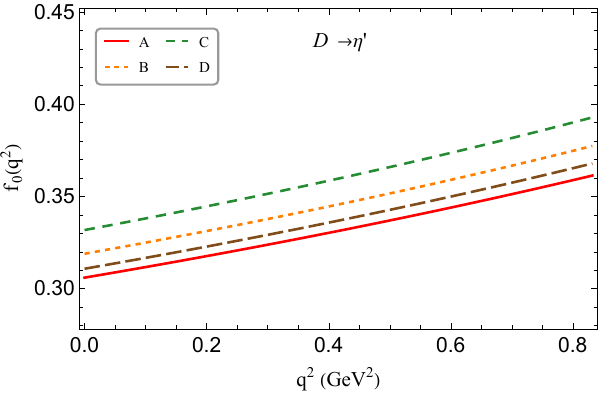} 
\end{subfigure}\non\vspace{4mm}
\begin{subfigure}[b]{0.49\textwidth}\centering
\includegraphics[width=\textwidth]{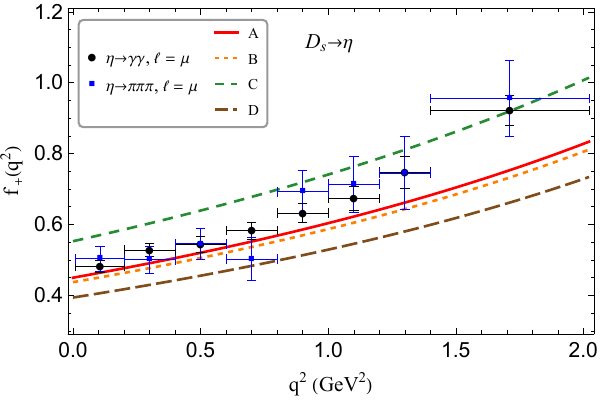} 
\end{subfigure}\hfill  
\begin{subfigure}[b]{0.49\textwidth}\centering
\includegraphics[width=\textwidth]{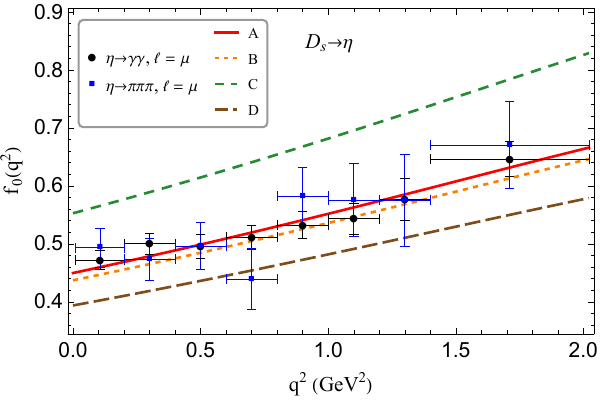} 
\end{subfigure}\non\vspace{4mm}
\begin{subfigure}[b]{0.49\textwidth}\centering
\includegraphics[width=\textwidth]{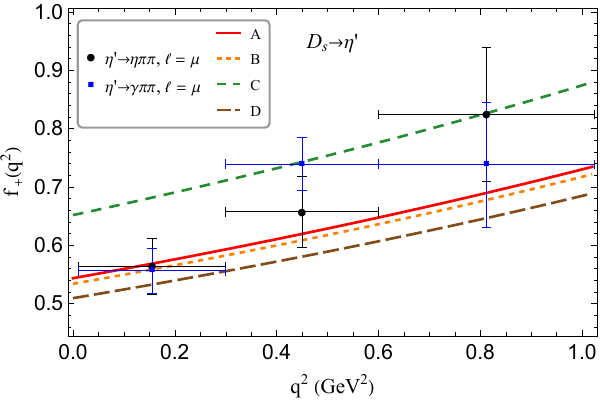}
\end{subfigure}\hfill  
\begin{subfigure}[b]{0.49\textwidth}\centering
\includegraphics[width=\textwidth]{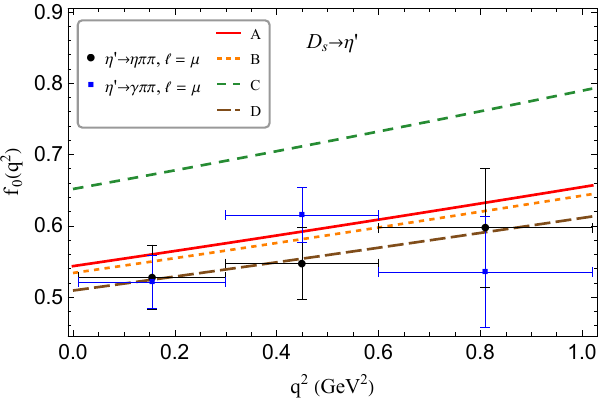} 
\end{subfigure}
\caption{LCSR predictions of the $D,D_{s}\to \eta^{(\prime)}$ FFs, $f_+(q^2)$ and $f_0(q^2)$, in the whole kinematical region, derived under four $\eta-\eta^\prime$ mixing parameter sets (A, B, C and D). For comparison, the BESIII measurements~\cite{BESIII:2025hjc,BESIII:2023gbn,BESIII:2024njj} are shown, which are represented by the black dots and blue squares with error bars. } 
\label{fig:ffs-mixing-schemes}
\end{figure}

\begin{figure}[t] 
\begin{center}
\begin{subfigure}[b]{0.49\textwidth} \centering
\includegraphics[width=\textwidth]{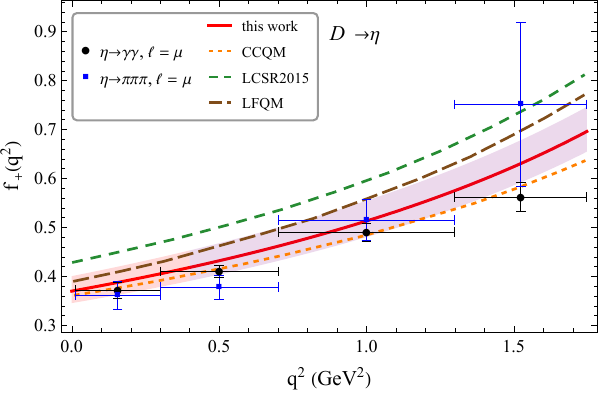}\end{subfigure} \hfill
\begin{subfigure}[b]{0.49\textwidth} \centering
\includegraphics[width=\textwidth]{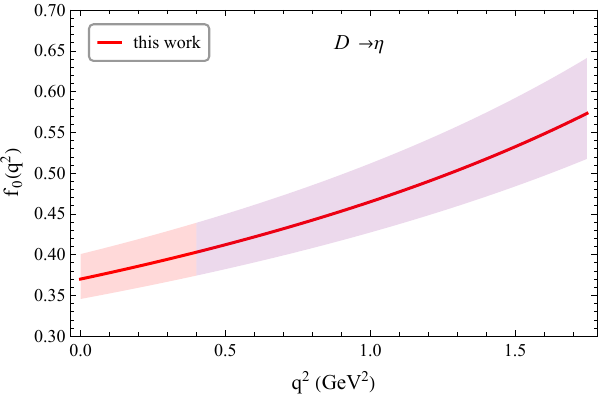}
\end{subfigure}\non\vspace{4mm}
\begin{subfigure}[b]{0.49\textwidth} \centering
\includegraphics[width=\textwidth]{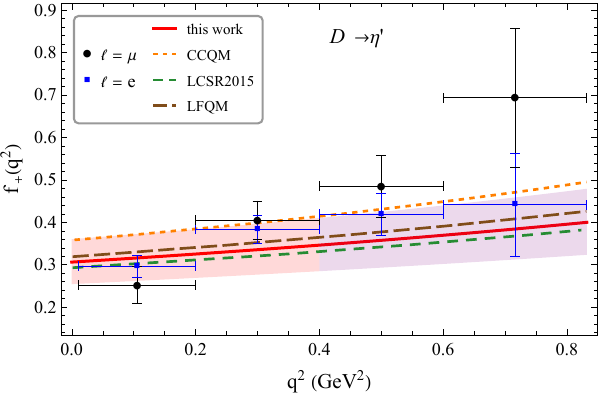} \end{subfigure} \hfill
\begin{subfigure}[b]{0.49\textwidth} \centering
\includegraphics[width=\textwidth]{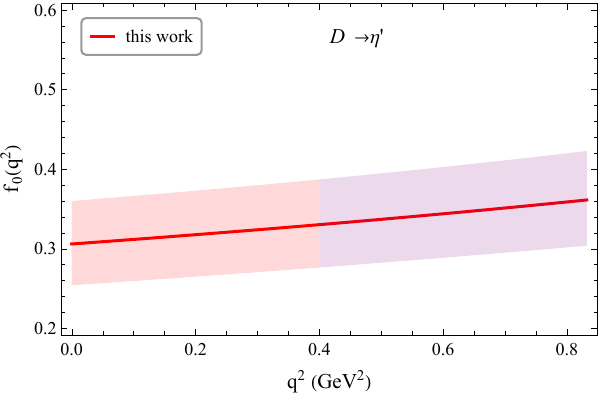}  
\end{subfigure}\non\vspace{4mm}
\begin{subfigure}[b]{0.49\textwidth} \centering
\includegraphics[width=\textwidth]{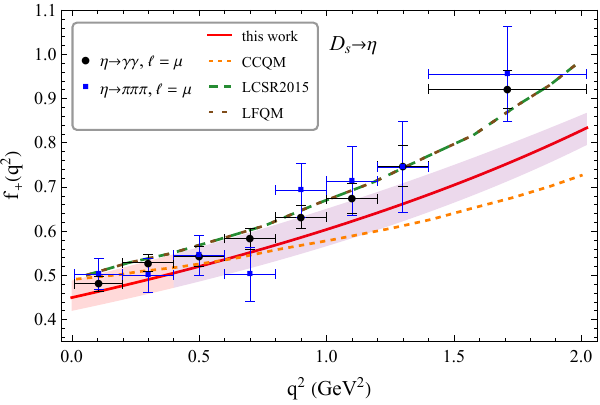}
\end{subfigure} \hfill  
\begin{subfigure}[b]{0.49\textwidth} \centering
\includegraphics[width=\textwidth]{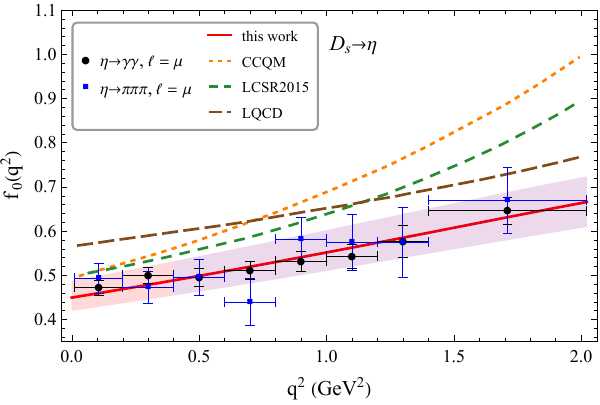}   \end{subfigure}\non\vspace{4mm}
\begin{subfigure}[b]{0.49\textwidth} \centering
\includegraphics[width=\textwidth]{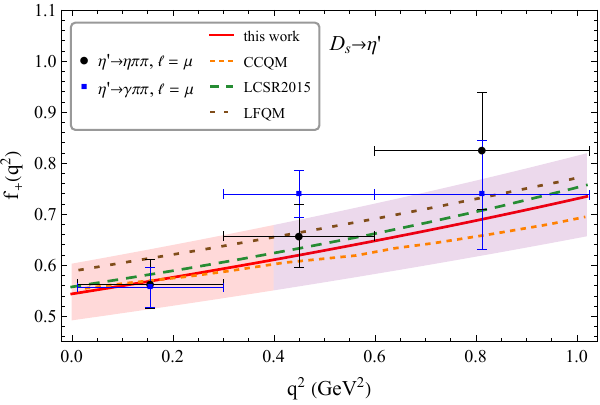}
\end{subfigure} \hfill
\begin{subfigure}[b]{0.49\textwidth} \centering
\includegraphics[width=\textwidth]{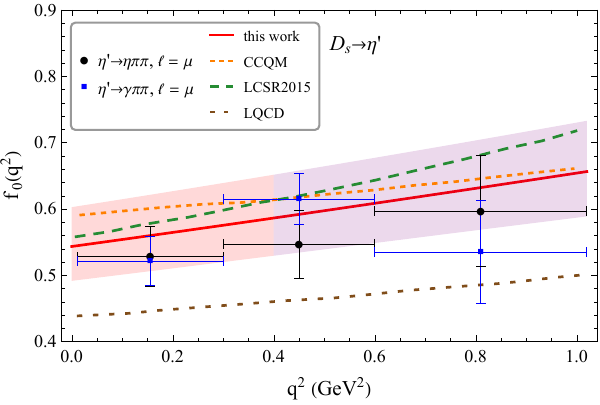} 
\end{subfigure}
\end{center}
\caption{LCSR predictions of the $D, D_{s} \to \eta^{(\prime)}$ FFs, derived with Set A mixing parameters. Alongside our results, we show for comparison the BESIII measurements~\cite{BESIII:2025hjc,BESIII:2023gbn,BESIII:2024njj} and other theoretical determinations from the CCQM~\cite{Ivanov:2019nqd}, LFQM~\cite{Verma:2011yw}, and LCSR2015~\cite{Duplancic:2015zna}. } \label{fig:DDs2etaetap-ffs}  
\end{figure}

Our LCSR predictions for the $D, D_s \to \eta^{(\prime)}$ transition FFs are shown across the full kinematical range in Fig.~\ref{fig:ffs-mixing-schemes}. Recent experimental extractions by BESIII collaboration~\cite{BESIII:2025hjc,BESIII:2023gbn,BESIII:2024njj} are also displayed for comparison, which are measured via the tagging channels $\eta \to \gamma\gamma, \eta \to \pi\pi\pi$ and $\eta^\prime \to \eta \pi\pi, \gamma \pi\pi$, with the lepton $\ell=\mu$ and $e$. With the LCDA parameters held fixed at their Set I values, we compute the FFs for all four mixing-parameter sets listed in Table~\ref{tab:parameter-mixing-t2LCDAs}.  We find that the Set A parameters lead to the most consistent interpretation of the BESIII results. This finding indicates that the experimental data are better described by a mixing scenario involving smaller decay constants and a larger mixing angle. In Fig.~\ref{fig:DDs2etaetap-ffs}, we further compare our results with previous LCSRs calculation~\cite{Duplancic:2015zna} and other theoretical determinations using various approaches, such as covariant constituent quark model (CCQM)~\cite{Ivanov:2019nqd}, light-front quark model (LFQM)~\cite{Verma:2011yw}, and lattice QCD (LQCD)~\cite{Bali:2014pva}. Our results for the low-$q^2$ region are shown as light magenta curves with bands, and the BCL extrapolations to higher $q^2$ as light purple ones. Good agreement is observed among the various approaches for the vector FF $f_+(q^2)$. In contrast, the LCSR predictions for the scalar FF $f_0(q^2)$ show a sizable discrepancy with other results, especially at high $q^2$ of small recoil.

\begin{figure}[t]\centering  
\begin{subfigure}[b]{0.49\textwidth}\centering
\includegraphics[width=\textwidth]{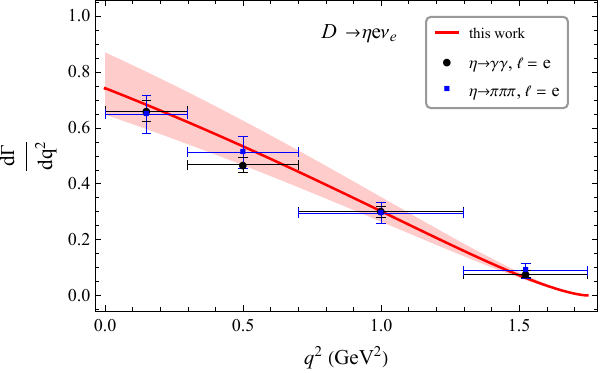}  
\end{subfigure}\hfill  
\begin{subfigure}[b]{0.49\textwidth}\centering
\includegraphics[width=\textwidth]{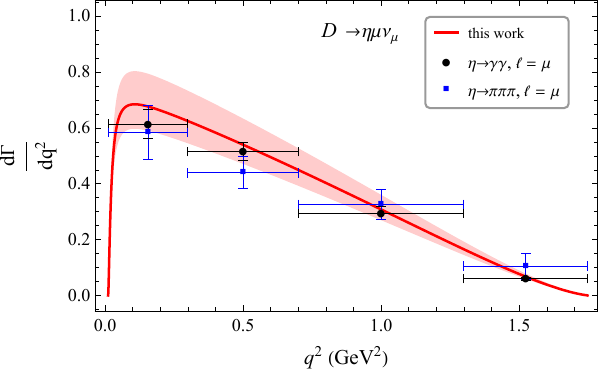} 
\end{subfigure} \non\vspace{4mm}
\begin{subfigure}[b]{0.49\textwidth}\centering
\includegraphics[width=\textwidth]{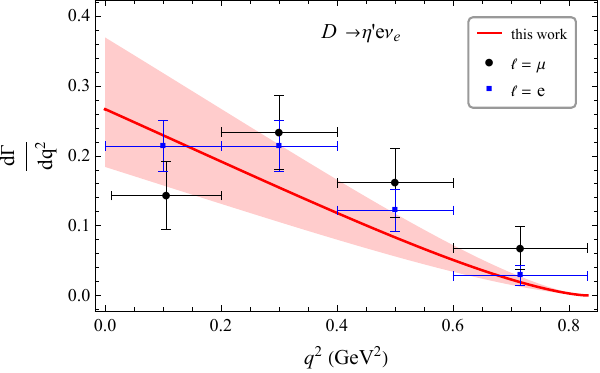} 
\end{subfigure}\hfill  
\begin{subfigure}[b]{0.49\textwidth}\centering
\includegraphics[width=\textwidth]{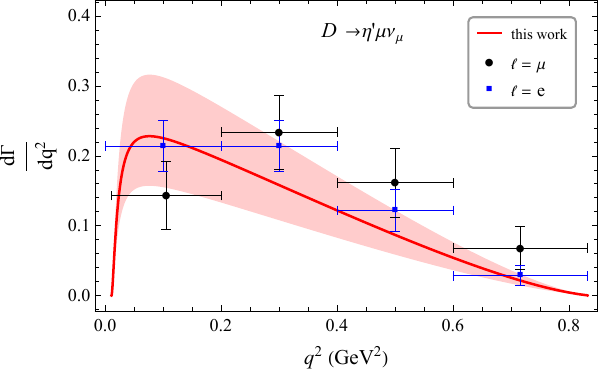}
\end{subfigure} \non\vspace{4mm}
\begin{subfigure}[b]{0.49\textwidth}\centering
\includegraphics[width=\textwidth]{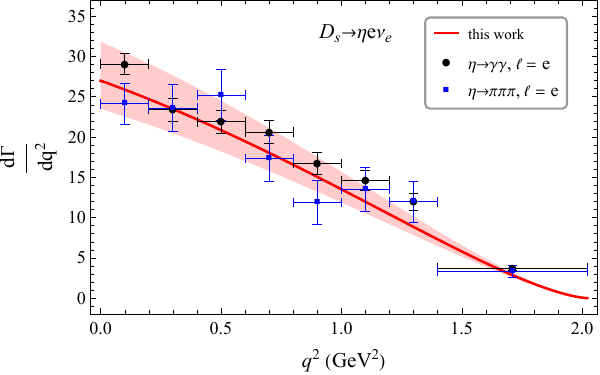}
\end{subfigure}\hfill 
\begin{subfigure}[b]{0.49\textwidth}\centering
\includegraphics[width=\textwidth]{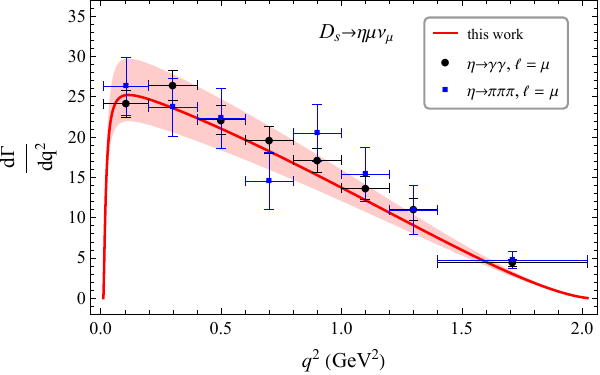} 
\end{subfigure} \non\vspace{4mm}
\begin{subfigure}[b]{0.49\textwidth}\centering
\includegraphics[width=\textwidth]{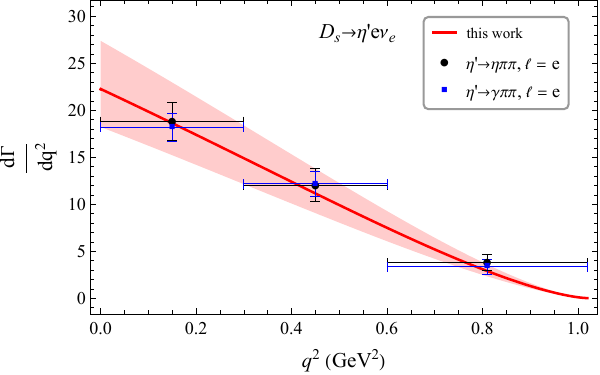} 
\end{subfigure}\hfill  
\begin{subfigure}[b]{0.49\textwidth}\centering
\includegraphics[width=\textwidth]{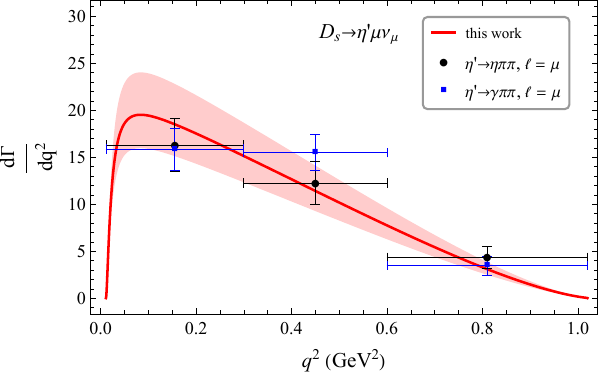}  
\end{subfigure}
\caption{LCSR predictions of differential decay rates
(in units of $10^{-15}\,\mathrm{GeV}^{-1}$),
derived using the Set A mixing parameters,
for the decays $D_{(s)}\to \eta^{(\prime)}\ell^+\nu_\ell$.
Left panels show the electronic modes with $\ell=e$,
while right panels correspond to the muonic modes with $\ell=\mu$.}
\label{fig:dGamma}
\end{figure}

\subsection{Differential decay rates and branching ratios}

The differential decay rate of the semileptonic decay $D \to \eta^{(\prime)} \ell^+ \nu_\ell$ takes the form~\cite{Faustov:2019mqr}
\beq 
\frac{d\Gamma}{dq^2}\bigg|_{D \to \eta^{(\prime)} \ell^+ \nu_\ell}
&=& \frac{G_F^2 \vert V_{cd} \vert^2}{24\pi^3} \frac{(q^2 - m_\ell^2)^2  p_{\eta^{(\prime)}} }{q^4 m_D^2} \Bigg[  m_D^2  p_{\eta^{(\prime)}}^2 \vert f_+(q^2)\vert^2 \non
&+& \frac{m_\ell^2}{2q^2} \left( 
m_D^2  p_{\eta^{(\prime)}}^2 \vert f_+(q^2)\vert^2 + \frac{3}{4}(m_D^2 - m_{\eta^{(\prime)}}^2)^2 \vert f_0(q^2) \vert^2 \right) \Bigg]\ , 
\label{eq:diff-decayrate} 
\eeq
where the scalar FF is suppressed by the lepton mass ${\cal O}(m_\ell^2/q^2)$. The Fermi coupling constant is taken as $G_F=1.166 \times 10^{-5}$ GeV$^{-2}$~\citep{ParticleDataGroup:2024cfk}. In the $D$-meson rest frame, the three-momentum modulus of the final meson is given by $p_{\eta^{(\prime)}} = \lambda^{1/2}(m_D^2, m_{\eta^{(\prime)}}^2, q^2)/(2m_D)$, where $\lambda(a,b,c) = a^2 + b^2 + c^2 - 2(ab + bc + ca)$ is the K\"all\'en function. For the $D_s \to \eta^{(\prime)} \ell \nu$ decays, the corresponding formulae are obtained by replacing the mass $m_D \to m_{D_s}$ and the CKM matrix element $ V_{cd} \to V_{cs} $, and by adapting the form factors appropriately.

Results of the differential decay rates for $D,D_{s} \to \eta^{(\prime)} \ell^+ \nu_{\ell}$ with respect to the momentum transfer squared $q^2$ are presented in Fig.~\ref{fig:dGamma}, where the left and right panels correspond to the $\ell = e$ and $\mu$ modes, respectively. Our theoretical predictions,  based on Set-A mixing parameters, are confronted with experimental data by BESIII~\cite{BESIII:2025hjc,BESIII:2023gbn,BESIII:2024njj}.
Good agreement is observed for the decays $D \to \eta \ell \nu_\ell$, $D_s \to \eta \ell \nu_\ell$, and $D_s \to \eta^\prime \ell \nu_\ell$. However, for $D \to \eta^\prime \ell \nu_\ell$, our predictions fall below the data across the intermediate and high $q^2$ region, particularly in the range $0.2 \leqslant q^2 \leqslant 0.6~\mathrm{GeV}^2$. This discrepancy, which was also observed in Ref.~\cite{BESIII:2024njj}, can be attributed to differences in the corresponding FFs between theoretical predictions and data-driven extractions, as shown in Fig.~\ref{fig:DDs2etaetap-ffs}. The discrepancy in the $D \to \eta^{(\prime)}$ FFs could be clarified in the future through more precise experimental data from BESIII, which would also be valuable to constrain the gluonic content of the $\eta$ and $\eta^\prime$ states~\cite{Escribano:2007cd,Cheng:2008ss,Fan:2012kn}.

\begin{table}[t]\centering
\caption{Branching fractions (in unit of $10^{-3}$) of the semileptonic decays $D, D_{s}^+\to \eta^{(\prime)} \ell^+ \nu_\ell$ , together with the PDG averages~\cite{ParticleDataGroup:2024cfk}, the BESIII measurements~\cite{BESIII:2023gbn,BESIII:2024njj,BESIII:2025hjc} and the recent LCSR determination~\cite{Hu:2025ool}.} 
\label{tab:Brs}\setlength{\tabcolsep}{12pt}
\begin{tabular}{l c c c c}\toprule
& $D^+\to \eta e^+ \nu_e$ & $D^+\to \eta \mu^+\nu_{\mu}$ & $D^+\to \eta^{\prime} e^+ \nu_e$& $D^+\to \eta^{\prime} \mu^+\nu_{\mu}$ \\
\midrule 
This work & $0.99^{+0.17}_{-0.13}$ & $0.97^{+0.17}_{-0.12}$ & $0.16^{+0.06}_{-0.05}$ & $0.15^{+0.06}_{-0.05}$ \\
PDG \cite{ParticleDataGroup:2024cfk} & $1.11\pm 0.07$ & $1.04\pm0.11$ & $0.20\pm 0.04$ & --- \\
BESIII~\cite{BESIII:2025hjc} & $0.98^{+0.29+0.28}_{-0.29-0.28}$ & $0.91^{+0.35+0.23}_{-0.35-0.23}$ & --- & --- \\
BESIII \cite{BESIII:2024njj} & --- & --- & $0.18^{+0.02+0.01}_{-0.02-0.01}$ & $0.19^{+0.03+0.01}_{-0.03-0.01}$ \\
\midrule  & $D_s^+\to \eta e^+ \nu_e$ & $D_s^+\to \eta \mu^+\nu_{\mu}$ & $D_s^+\to \eta^{\prime} e^+ \nu_e$& $D_s^+\to \eta^{\prime} \mu^+\nu_{\mu}$ \\ \midrule
This work & $20.35^{+3.41}_{-2.52}$ & $19.96^{+3.35}_{-2.48}$ & $7.80^{+1.83}_{-1.45}$ & $7.45^{+1.75}_{-1.38}$ \\
LCSR \cite{Hu:2025ool} & $23.46^{+4.18}_{-3.31}$ & $23.20^{+4.13}_{-3.27}$ & $7.92^{+1.41}_{-1.18}$ & $7.73^{+1.38}_{-1.15}$ \\
PDG \cite{ParticleDataGroup:2024cfk} & $22.70\pm 0.60$ & $22.40 \pm 0.70$ & $8.10 \pm 0.60$ & $8.00 \pm 0.60$ \\
BESIII \cite{BESIII:2023gbn} & --- & $22.35^{+0.51+0.52}_{-0.51-0.52}$ & --- & $8.01^{+0.55+0.28}_{-0.55-0.28}$ \\
\bottomrule \end{tabular}\end{table}

The partial decay width can be obtained by integrating the differential decay rates in Eq.~\eqref{eq:diff-decayrate} over the entire kinematical range of momentum transfer squared, namely,
\beq 
\Gamma(D_{(s)} \to \eta^{(\prime)} \ell^+ \nu_\ell) = \int_{m_{\ell}^2}^{ (m_{D_{(s)}}-m_{\eta^{(\prime)}})^2}{\rm d}q^2 \bigg[\frac{{\rm d}\Gamma}{{\rm d}q^2}\bigg|_{D_{(s)} \to \eta^{(\prime)} \ell^+ \nu_\ell}\bigg]\ ,  
\label{eq:decaywidth}
\eeq
and the branching fraction is defined by ${\cal B} = \Gamma(D_{(s)} \to \eta^{(\prime)} \ell^+ \nu_\ell) \cdot \tau_{D_{(s)}} $. We use the PDG values $\tau_D=(1.033\pm 0.005)$ ps and $\tau_{D_s}=(0.5012\pm 0.0022)$ ps for the lifetimes of $D_{(s)}$ mesons. Our predictions are presented in Table \ref{tab:Brs}, together with world averages~\cite{ParticleDataGroup:2024cfk}, recent BESIII measurements~\cite{BESIII:2023gbn,BESIII:2024njj,BESIII:2025hjc} and previous LCSR determination~\cite{Hu:2025ool}. For the decays $D_s \to \eta^{(\prime)} \ell^+ \nu_\ell$ induced by weak $c\to s$ currents, the experimental precision exceeds that of our LCSR calculations by roughly a factor of 4. In contrast, for the $D \to \eta^{(\prime)} \ell^+ \nu_\ell$ decays induced by weak $c\to d$ currents, the LCSR predictions remain more precise than the BESIII results and are comparable to the PDG averages. 
Notably, the aforementioned discrepancy in the $D \to \eta^\prime$ differential decay rate does not manifest in the corresponding branching fraction. This is partly due to phase-space suppression in the high-$q^2$ region when integrating over $q^2$ in Eq. (\ref{eq:decaywidth}). As illustrated in Fig. \ref{fig:dGamma}, the larger LCSR results in the low-$q^2$ region compensate for the lower curve at intermediate-to-large $q^2$.  

\section{Summary and outlook}\label{sec:conclu}

Inspired by recent BESIII measurements of $D, D_s \to \eta^{(\prime)} \ell^+ \nu_\ell$ decays~\cite{BESIII:2023gbn,BESIII:2024njj,BESIII:2025hjc}, we revisit the relevant transition FFs using QCD LCSRs approach. Our calculation demonstrates good convergence of the OPE series, beginning with the twist-3 contribution. The obtained FFs are overwhelmingly dominated by the two-particle twist-3 contribution, which benefits from chiral enhancement due to the Goldstone-boson nature of the final $\eta^{(\prime)}$ mesons. Owing to destructive interference between the leading-twist and two-particle twist-3 components, the NLO hard-gluon correction reduces the FFs by 2\%–3\%. In addition, soft corrections from three-particle LCDAs and two-gluon configurations are negligible. This smallness continues for higher-twist (5 and 6) contributions, which we estimate via a factorization hypothesis into local condensates and leading-twist LCDAs. 

High-accuracy LCSR FFs provide robust inputs to scrutinize the $\eta-\eta^\prime$ mixing mechanism in the semileptonic decays $D, D_s \to \eta^{(\prime)} \ell^+ \nu_\ell$. The recent BESIII data strongly favors a mixing-parameter set characterized by relatively small decay constants $f_{\eta_q} = \left( 1.02^{+0.02}_{-0.05} \right) f_\pi$ and $f_{\eta_s} = \left( 1.37^{+0.04}_{-0.06} \right) f_\pi$, as well as a larger mixing angle $\phi = 39.6^{+1.2}_{-2.1}$ degree. It is found that, for the $D \to \eta^\prime \ell^+ \nu_\ell$ decay, the predicted differential decay rate shows a slight tension with the BESIII result, particularly in the intermediate-to-high momentum transfer region. Although the current theoretical precision is comparable to the experimental uncertainty, further refined measurements and more precise FF determinations remain essential to unravel the potential influence of gluonic components in the $D_{(s)} \to \eta^{(\prime)} \ell^+ \nu_\ell$ decays. 

\acknowledgments
We are grateful to Tao Luo, Xiang Pan, Yu-Ming Wang and Shu-Lei Zhang for fruitful discussions. This work is supported by the National Key R$\&$D Program of China under Contracts No. 2023YFA1606000; by the National Science Foundation of China (NSFC) under Contract No. 12575098, No.~12275076 and No.~12335002; by the Science Fund for Distinguished Young Scholars of Hunan Province under Grant No. 2024JJ2007; by the Fundamental Research Funds for the Central Universities under Contract No. 531118010379.

\appendix

\section{Explicit expressions of the $D \to \eta_{q,g}$ form factors} \label{app:FF-piece}

This appendix details the LO results for the $D \to \eta_q$ transition FFs, which are categorized by twist. Additionally, the $D \to \eta_{g}$ FFs at NLO is provided. The lengthier NLO corrections to the $D \to \eta$ FFs are not reproduced here, whose complete expressions can be found in Ref.~\cite{Duplancic:2008ix}. The result for the $D_s \to \eta_s$ transition can be derived from that of $D \to \eta_q$ by directly substituting the corresponding parameters in the LCSRs and LCDAs.

The twist-2, twist-3 and twist-4 contributions to the $D \to \eta_q$ transition FFs read 
\beq
&&f_{+, {\rm LO}}^{D \to \eta_q, (t=2)}(q^2) = \frac{e^{m_D^2/M^2}}{2 m_D^2 f_D^2} \, m_c^2 \, \int_{u_0}^1\frac{du}{u}e^{-\frac{s(u)}{M^2}} \,\phi_2(u)\ , \\ 
&&f_{+, {\rm LO}}^{D \to \eta_q, (t=3)}(q^2) = \frac{e^{m_D^2/M^2}}{2 m_D^2 f_D^2} \int_{u_0}^1 du e^{-\frac{s(u)}{M^2}} \Big\{ 
m_c \mu_{\eta_q} \Big( \phi_{3}^{p}(u) 
+ \frac{1}{3} \frac{\phi_{3}^{\sigma}}{u} 
- \frac{m_c^2+q^2-u^2m_{\eta_q}^2}{\triangle(u,q^2)} \frac{d\phi_{3}^{\sigma}(u)}{du}\non
&&\hspace{1.5cm}  + \frac{4 u m_{\eta_q}^2 m_c^2}{\left[ \triangle(u,q^2) \right]^2}\phi_{3}^{\sigma}(u)) \Big)
- m_c \frac{f_{3 \eta_q}}{f_{\eta_q}} 
\Big( \frac{1}{\triangle(u,q^2)} \frac{dI_{3}}{du}-\frac{2 u m_{\eta_q}^2}{ \left[ \triangle(u,q^2) \right]^2} I_{3} \Big) \Big\}\ , \\
&&f_{+, {\rm LO}}^{D \to \eta_q, (t=4)}(q^2) = \frac{e^{m_D^2/M^2}}{2 m_D^2 f_D^2} 
\int_{u_0}^1 du e^{-\frac{s(u)}{M^2}} \Big\{ 
\frac{u m_c^2}{\triangle(u,q^2)} \psi_{4}(u)
+ \left( \frac{m_c^2}{\triangle(u,q^2)} 
- \frac{2 u^2 m_{\eta_q}^2 m_c^2}{\left[ \triangle(u,q^2) \right]^2} \right) \int_0^u dv\psi_{4}(v) \non
&&\hspace{1.5cm} - \frac{u m_c^4}{4 \left[ \triangle(u,q^2) \right]^2} \frac{d^2\phi_{4}(u)}{du^2} 
+ \frac{3 u^2 m_{\eta_q}^2 m_c^4}{2 \left[ \triangle(u,q^2) \right]^3}\frac{d\phi_{4}(u)}{du} 
- \frac{3 u^3 m_{\eta_q}^4 m_c^4 }{\left[ \triangle(u,q^2) \right]^4} \phi_{4}(u) \non
&&\hspace{1.5cm} -\frac{m_c^2}{\triangle(u,q^2)} \frac{dI_{4}}{du}
+ \frac{2 u m_{\eta_q}^2 m_c^2}{\left[ \triangle(u,q^2) \right]^2} I_{4} 
- \left( \frac{2 u m_{\eta_q}^2 m_c^2}{\left[ \triangle(u,q^2) \right]^2} + \frac{2 u^2 m_{\eta_q}^2 m_c^2}{\left[ \triangle(u,q^2) \right]^2} \frac{d}{du} 
- \frac{8 u^3 m_{\eta_q}^4 m_c^2}{\left[ \triangle(u,q^2) \right]^3} \right) \bar{I}_{4}(u) \non 
 &&\hspace{1.5cm} + \left( \frac{2 u m_c^2 m_{\eta_q}^2 (m_c^2-q^2-u^2m_{\eta_q}^2)}{\left[ \triangle(u,q^2) \right]^3}\frac{d}{du} - \frac{12 u^2 m_c^2 m_{\eta_q}^4 (m_c^2-q^2-u^2m_{\eta_q}^2)}{\left[ \triangle(u,q^2) \right]^4} \right) \int_u^1d\xi \bar{I}_{4}(\xi) 
\Big\}\ . 
\label{eq:D2etaq-Fp-2p3p-t234-LO} \eeq 
Here, $u$ is the anti-quark's longitudinal momentum fraction, the integration lower limit $u_0$ is determined by solving $s(u,q^2) = s_0$, where $s(u,q^2) = \left(m_c^2-q^2\bar{u}+m_{\eta_q}^2 u\bar{u} \right)/u$. To simplify the formalism, we introduce the kinematical function $\triangle(u,q^2) = m_c^2-q^2+u^2m_{\eta_q}^2$. 
$\phi_2$, $\phi_3^{p, \sigma}$ and $\psi_4, \phi_4$ are the twist-2, twist-3 and twist-4 LCDAs defined for the lowest Fock state ($q{\bar q}$). The contributions from three-particle Fock state ($q{\bar q}g$) are absorbed into the auxiliary functions: 
\beq 
&&I_{3}(u) = \int_0^u d\alpha_1 \int_{\frac{u-\alpha_1}{1-\alpha_1}}^1 \frac{dv}{v} \left[3(2v-1)um_{\eta_q}^2 + 4v(p\cdot q)\right]  \Phi_{3}(\alpha_i)\Big\vert_{\substack{\alpha_1=1-\alpha_2-\alpha_3 \\ \alpha_3=(u-\alpha_1)/v}}\ ,\\
&&I_{4}(u)=\int_0^u d\alpha_1 \int_{\frac{u-\alpha_1}{1-\alpha_1}}^1 \frac{dv}{v}[2\Psi_{4}(\alpha_i)-\Phi_{4}(\alpha_i)+2\tilde{\Psi}_{4}(\alpha_i)- \tilde{\Phi}_{4}(\alpha_i)] \Big\vert_{\substack{\alpha_1=1-\alpha_2-\alpha_3 \\ \alpha_3=(u-\alpha_1)/v}}\ ,\\
&&\bar{I}_{4}(u)=\int_0^u d\alpha_1 \int_{\frac{u-\alpha_1}{1-\alpha_1}}^1 \frac{dv}{v}[\Psi_{4}(\alpha_i)+\Phi_{4}(\alpha_i)+\tilde{\Psi}_{4}(\alpha_i)+\tilde{\Phi}_{4}(\alpha_i)]\Big\vert_{\substack{\alpha_1=1-\alpha_2-\alpha_3 \\ \alpha_3=(u-\alpha_1)/v}}\ ,
\label{eq:D2etaq-Fp-3p-t34-LO}
\eeq
where $\Psi_{3}$ and $\Psi_{4}, \Phi_{4}, {\tilde \Psi}_4, {\tilde \Phi}_4$ are twist-3 and twist-4 three particle LCDAs, respectively. 

In Ref. \cite{Rusov:2017chr}, the twist-5 and twist-6 contributions to the $D \to \eta_q$ FFs are expressed as a convolution of lower-twist LCDAs with quark vacuum condensates. 
\beq
f_{+, {\rm LO}}^{D \to \eta_q, (t=5,6)}(q^2) 
&=& \frac{e^{m_D^2/M^2}}{2 m_D^2 f_D^2} \langle {\bar q}q \rangle \frac{\alpha_s \pi C_F m_c}{N_C} \non 
&~& \int_{m_c^2}^{\infty} ds \sum_{n=2,3,4} \frac{(-1)^{n-1}}{(n-1)!} g_n(q^2,s) 
\times\frac{d^{n-1}}{ds^{n-1}} \theta(s_0-s)e^{-\frac{s}{M^2}}\ . \label{eq:D2etaq-Fp-2p-t56-LO}
\eeq
The additional functions $g_n(q^2, s)$, which depend on both the momentum transfer and internal virtuality, are given by 
\beq &&g_2(q^2,s)=\frac{1}{m_c^2-q^2} \int_{u_1}^1\frac{du}{u} \, \phi_2( \frac{u_1}{u}) + \frac{2}{m_c^2-q^2} \int_{u_2}^1 \frac{du}{u}(1-2 u \bar{u}) \, \phi_2(\frac{u_2}{u})\ , \\
&&g_3(q^2,s)=\frac{4q^2(s-q^2)}{(m_c^2-q^2)^2} \int_{u_1}^1 du \bar{u} \, \phi_2(\frac{u_1}{u}) -\frac{4\mu_{\eta_q}m_c}{m_c^2-q^2} \int_{u_1}^1 du \bar{u} \, \phi_{3}^{p}(\frac{u_1}{u}) \non
&& \hspace{1.5cm} +\frac{4m_c^2(s-q^2)}{(m_c^2-q^2)^2} \int_{u_2}^1 du \bar{u} \, \phi_2(\frac{u_2}{u}) + \frac{4\mu_{\eta_q}m_c}{m_c^2-q^2} \int_{u_2}^1 du \bar{u} \, \phi_{3}^{p}(\frac{u_2}{u})\ , \\
&&g_4(q^2,s) = 2\mu_{\eta_q} m_c \frac{(s-q^2)^2(m_c^2+q^2)}{(m_c^2-q^2)^3} \left[  \int_{u_1}^1 du u\bar{u} \, \phi_{3}^{\sigma}(\frac{u_1}{u}) + \int_{u_2}^1 du u\bar{u} \, \phi_{3}^{\sigma}(\frac{u_2}{u}) \right]\ .
\eeq
The lower limits in the above integrals are 
$u_1 \equiv u_1(s,q^2) = (m_c^2-q^2)/(s-q^2)$ and $u_2  \equiv u_2(s,q^2) = (s-m_c^2)(s-q^2)$.

The expression for the NLO gluonic contribution has been derived in Ref.~\cite{Duplancic:2015zna}, which are
\beq &&f_{+, {\rm NLO}}^{D \to \eta_g, (t=2)}(q^2) =  \frac{e^{m_D^2/M^2}}{2 m_D^2 f_D^2} \frac{\alpha_s}{4\pi} \, b_{2}^{g}\int_{m_c^2}^{s_0} ds \, e^{-s/M^2}f_+^{gg}(s,q^2)\ , \\
&&f_+^{gg}(s,q^2) = 20 m_c^2 \frac{s-m_c^2}{27\sqrt{3}(s-q^2)^5} \Big\{3(m_c^2-q^2) \left[ 5m_c^4-5m_c^2(q^2+s)+q^4+3q^2s+s \right] \non
&&\hspace{1.5cm} \times \left[ 2\ln (\frac{s-m_c^2}{m_c^2}) - \ln (\frac{\mu^2}{m_c^2}) \right] 
- (37 m_c^6-m_c^4(56q^2+55s) \non
&&\hspace{1.5cm} + m_c^2(18q^4+76q^2s+17s^2)+3q^6-27q^4s-11q^2s^2-2s^3) \Big\}\ . 
\eeq

For the combined form factor $f_{+-, {\rm LO}}^{D \to \eta_q}(q^2) \equiv f_{+, {\rm LO}}^{D \to \eta_q}(q^2) + f_{-, {\rm LO}}^{D \to \eta_q}(q^2)$, the twist-2 LCDA does not contribute and the current calculation include the twist-3 and twist-4 terms at LO, which read 
\beq &&f_{+-, {\rm LO}}^{D \to \eta_q, (t=3)}(q^2) = \frac{e^{m_D^2/M^2}}{m_D^2 f_D^2} 
\int_{u_0}^1 du \, e^{-\frac{s(u,q^2)}{M^2}} 
\Big\{ \frac{m_c\mu_{\eta_q}}{u} \, \phi_{3}^{p}(u) + \frac{m_c\mu_{\eta_q}}{6u} \frac{d\phi_{3}^{\sigma}}{du} \non
&&\hspace{1.5cm} + \frac{f_{3\eta_q}}{f_{\eta_q}} m_c m_{\eta_q}^2 
\left[ \frac{1}{\triangle(u,q^2)} \frac{d\tilde{I}_{3}(u)}{du} - \frac{2u m_{\eta_q}^2}{\left[\triangle(u,q^2) \right]^2} \tilde{I}_{3}(u) \right] \Big\}\ , \\
&&f_{+-, {\rm LO}}^{D \to \eta_q, (t=4)}(q^2) = \frac{e^{m_D^2/M^2}}{m_D^2 f_D^2} 
\int_{u_0}^1 du \, e^{-\frac{s(u,q^2)}{M^2}} 
\Big\{ \frac{m_c^2}{\triangle(u,q^2)} \, \psi_{4}(u) - \frac{2 u m_{\eta_q}^2 m_c^2}{\left[ \triangle(u,q^2) \right]^2} \int_0^u dv \, \psi_{4}(v) \non 
&&\hspace{1.5cm} + \frac{2u m_{\eta_q}^2 m_c^2}{\left[ \triangle(u,q^2) \right]^2} \left( \frac{d^2}{du^2} -\frac{6 u m_{\eta_q}^2}{\triangle(u,q^2)} \frac{d}{du} + \frac{12 u^2 m_{\eta_q}^4}{\left[ \triangle(u,q^2) \right]^2} \right) 
\int_u^1 d\xi \, \bar{I}_{4}(\xi) \Big\}\ , 
\label{eq:D2etaq-Fpm-3p-t34-LO}
\eeq
where the newly introduced auxiliary function is given by
\beq \tilde{I}_{3} = \int_0^u d\alpha_1 \int_{\frac{u - \alpha_1}{1 - \alpha_1}}^1 
\frac{dv}{v} \left[ (3-2v) \right] 
\Phi_{3}(\alpha_i) \Big\vert_{\substack{\alpha_1 = 1 - \alpha_2 - \alpha_3\\ \alpha_3=(u-\alpha_1)/v}}\ .
\label{eq:D2etaq-Fpm-3p-t3-LO}\eeq

\section{LCDAs of light pseudoscalar mesons} \label{app:LCDAs}

The LCDAs of isoscalar pseudoscalar mesons ($\eta_r=\eta_{q,s}$) are defined via the corresponding bilocal matrix element with the approximate light-cone distance $x^2 \to 0$~\cite{Ball:2006wn}. 
At the lowest Fock state ($q{\bar q}$), it reads  
\beq
\langle{\eta_r(p)}|\bar{q}_{\omega}^i(x_1)q_{\xi}(x_2)|{0}\rangle 
&=& \frac{i\delta^{ij}}{12}f_{\eta_r}
\int_0^1 du e^{iup \cdot x_1+i\bar{u}p\cdot x_2} 
\Big\{ [\slashed{p}\gamma_5]_{\xi \omega} \phi(u) \non
&-& [\gamma_5]_{\xi \omega} \mu_{\eta_r}\phi_{3}^p(u) + \frac{1}{6}[\sigma_{\beta \tau}\gamma_5]_{\xi \omega}p_{\beta}(x_1-x_2)_{\tau}\mu_{\eta_r}\phi_{3}^{\sigma}(u) \non
&+& \frac{1}{16}[\slashed{p}\gamma_5]_{\xi \omega}(x_1-x_2)^2\phi_{4}(u) - \frac{i}{2}[(\slashed{x}_1-\slashed{x}_2)\gamma_5]_{\xi \omega}\int_0^u\psi_{4}(v)dv \Big\}\ .
\label{eq:LCDAs-qqbar}
\eeq
In the Fock state of quark-antiquark-gluon ($q{\bar q}g$), it is defined by 
\beq 
&~&\langle{\eta_r(p)}|\bar{q}_{\omega}^i(x_1) g_s G_{\mu\nu}^a(x_3) q_{\xi}^j(x_2)|{0}\rangle \non
&=& \frac{\lambda_{ji}^a}{32}\int \mathcal{D}\alpha_ie^{ip(\alpha_1x_1+\alpha_2x_2+\alpha_3x_3)}
\Big\{ if_{3\eta_r} [\sigma_{\lambda\rho}\gamma_5]_{\xi \omega}(p_{\mu}p_{\lambda}g_{\nu\rho}-p_{\nu}p_{\lambda}g_{\mu\rho})\Phi_{3}(\alpha_i) \non
&-& f_{\eta_r} [\gamma_{\lambda}\gamma_5]_{\xi \omega}\Big[ (p_{\nu}g_{\mu\lambda}-p_{\mu}g_{\nu\lambda})\Psi_{4}(\alpha_i)+\frac{p_{\lambda}(p_{\mu}x_{\nu}-p_{\nu}x_{\mu})}{(p\cdot x)}(\Phi_{4P}(\alpha_i)+\Psi_{4}(\alpha_i)) \Big] \non
&-& \frac{if_{\eta_r}}{2}\epsilon_{\mu\nu\delta\rho}[\gamma_{\lambda}]_{\xi\omega} \Big[ (p^{\rho}g^{\delta\lambda}-p^{\delta}g^{\rho\lambda})\tilde{\Psi}_{4}(\alpha_i)+\frac{p_{\lambda}(p^{\delta}x^{\rho}-p^{\rho}x^{\delta})}{(p\cdot x)}(\tilde{\Phi}_{4P}(\alpha_i)+\tilde{\Psi}_{4}(\alpha_i)) \Big] \Big\}\ , \label{eq:LCDAs-qqbarg} 
\eeq
with the integral measure $\mathcal{D}\alpha_i = d\alpha_1 d\alpha_2 d\alpha_3 \delta(1-\alpha_1-\alpha_2-\alpha_3)$. 
Furthermore, the LCDAs of two-gluon ($gg$) Fock state takes the form~\cite{Ball:2007hb} 
\beq \langle{\eta_g(p)}|G_{\mu x}(x)[x,-x] \tilde{G}^{\mu x}(-x)|{0}\rangle=f_{\eta^{(\prime)}}^1\frac{C_F}{2\sqrt{3}}(px)^2\int_0^1due^{-i(2u-1)px}\phi^g(u). \label{eq:LCDAs-gg} \eeq 

In Eqs.~\eqref{eq:LCDAs-qqbar}-\eqref{eq:LCDAs-gg},  $\phi(u)$ and $\phi^g$ are leading-twist LCDAs, $\phi_{3}^p(u), \phi_{3}^{\sigma}(u)$ and $\phi_{4}(u), \psi_{4}(v)$ are two particle twist-3 and twist-4 LCDAs, respectively. In addition, $\Phi_{3}(\alpha_i)$ is 
three particle twist-3 LCDAs, and $\Phi_{4}(\alpha_i), \Psi_{4}(\alpha_i), \tilde{\Phi}_{4}(\alpha_i), \tilde{\Psi}_{4}(\alpha_i)$ are twist-4 LCDAs. The derivation of their parameterizations is a use of conformal symmetry in the massless QCD \cite{Braun:2003rp}, which yields the expansion in terms of Gegenbauer polynomials \cite{Ball:1998je,Braun:1999uj,Bijnens:2002mg,Ball:2006wn}. We here quote the expressions for these LCDAs twist by twist: 
\beq \phi(u) &=& 6u(1-u) \Big[ 1+\sum_{n=2,4\cdots} a_n \, C_n^{3/2}(2u-1) \Big]\ , \non
\phi^g(u) &=& u^2(1-u)^2 \sum_{n=2,4...} b_n^{g} \, C_{n-1}^{5/2} 2u-1)\ ; \label{eq:LCDAs-t2} \\
\phi_{3}^{p}(u) &=& 1+30\eta_3 C_2^{1/2}(u-\bar{u})-3\eta_3 \omega_3 C_4^{1/2}(u-\bar{u})\ , \non
\phi_{3}^{\sigma}(u) &=& 6u(1-u)(1+5\eta_3 (1-\frac{w_{3}}{10})C_2^{3/2}(u-\bar{u}))\ , \non
\Phi_{3}(\alpha_i) &=& 360 \alpha_1 \alpha_2 \alpha_3^2 
\Big\{ 1 + \lambda_{3} (\alpha_1-\alpha_2) + \omega_{3} \frac{1}{2}(7\alpha_3-3) \Big\}\ ; \label{eq:LCDAs-t3} \\
\psi_{4}(u) &=& \frac{20}{3}\delta^2C_2^{1/2}(2u-1)\ , \non
\phi_{4}(u) &=& 
\frac{200}{3}\delta^2u^2\bar{u}^2+21\delta^2 \omega_4 \Big\{u\bar{u}(2+13u\bar{u})+2u^3(10-15u+6u^2)\ln u\non
&+&2\bar{u}^3(10-15\bar{u}+6\bar{u}^2)\ln\bar{u}\Big\}\ , \non
\Phi_{4}(\alpha_i) &=& 120\alpha_1\alpha_2\alpha_3 
\delta^2 \epsilon(\alpha_1-\alpha_2)\ , \non
\tilde{\Phi}_{4}(\alpha_i) &=& -120\alpha_1\alpha_2\alpha_3 \delta^2 \Big\{ \frac{1}{3}+\epsilon(1-3\alpha_3) \Big\}, \non
\Psi_{4}(\alpha_i) &=& 30\delta^2(\alpha_1-\alpha_2)\alpha_3^2\Big[\frac{1}{3}+2\epsilon(1-2\alpha_3)\Big]\ , \non
\tilde{\Psi}_{4}(\alpha_i) &=& 
30\delta^2\alpha_3^2(1-\alpha_3)\Big[\frac{1}{3}+2\epsilon(1-2\alpha_3)\Big]\ ,\label{eq:LCDAs-t4}
\eeq
where the abbreviations $\eta_{3}=f_{3\eta_r}/(f_{\eta_r} \mu_{\eta_r})$ and $\epsilon=21\omega_4/8$ have been used. 


%

\end{document}